\documentclass[
  aps,prx,reprint,superscriptaddress,
  amsmath,amssymb,longbibliography,
  nobalancelastpage
]{revtex4-2}

\usepackage{amsthm}
\usepackage[hidelinks]{hyperref}

\newtheorem{theorem}{Theorem}
\newtheorem{proposition}[theorem]{Proposition}

\begin{document}

\title{Sample-Based Quantum Linear Solver and Bounds on Krylov-State Sampling}

\author{Byeongyong Park}
\affiliation{Center for Quantum Information Processing, University of Seoul, 163 Seoulsiripdae-ro, Dongdaemun-gu, Seoul 02504, Republic of Korea}

\author{Keunhong Jeong}
\email[Contact author: ]{doas1mind@sogang.ac.kr}
\affiliation{Department of Chemistry, Sogang University, 35 Baekbeom-ro, Mapo-gu, Seoul 04107, Republic of Korea}

\author{Doyeol (David) Ahn}
\email[Contact author: ]{dahn@uos.ac.kr}
\affiliation{Center for Quantum Information Processing, University of Seoul, 163 Seoulsiripdae-ro,
Dongdaemun-gu, Seoul 02504, Republic of Korea}
\affiliation{Singularity Quantum Inc., 9506 Villa Isle Drive, Villa Park, California 92861, USA}

\begin{abstract}
To solve linear systems $Ax=b$, we propose a sample-based quantum linear solver combining quantum Krylov-state sampling with classical variational diagonalization.
Rather than simulating the cost Hamiltonian $H_b(A)=A^\dagger(I-|b\rangle\langle b|)A$, we measure quantum Krylov states $e^{-iAT}|b\rangle$ generated by the original Hermitian operator $A$ in the computational basis.
We then classically diagonalize the projected cost Hamiltonian to reconstruct an approximate solution.
Using a representation of $A^{-1}|b\rangle$ as a linear combination of quantum Krylov states, we derive lower bounds on computational-basis sampling probabilities from solution amplitudes.
We establish support-discovery guarantees for finite nonnegative-time grid sampling and signed-time sampling, which permits both positive and negative evolution times.
The former bounds the maximum evolution time, whereas the latter improves ideal shot complexity at the cost of occasional long evolutions.
For a condition-number bound $\kappa$ on $A$, a lower bound $\beta_L$ on the solution probability weight of each of $L$ target basis states, and failure probability $\delta$, signed-time sampling achieves $O(\kappa^2\beta_L^{-1}\log(L/\delta))$ shots, matching a worst-case lower bound up to constants within an evolution-only observation model in a specified parameter regime.
Finally, we analyze the effects of finite-length qDRIFT compilation and circuit reuse on sampling and derive an infidelity bound for the reconstructed solution.
\end{abstract}

\maketitle

\section{Introduction}\label{sec:introduction}
Solving linear systems of equations, $Ax=b$, is a fundamental computational task in scientific computing, optimization, and the numerical solution of differential equations.
However, as the system dimension or the condition number of the matrix increases, obtaining accurate solutions can become computationally demanding~\cite{saad}. 
This challenge has motivated research into quantum algorithms for solving linear systems.
In particular, the Harrow--Hassidim--Lloyd (HHL) algorithm demonstrated favorable theoretical complexity for preparing a quantum state proportional to the solution under suitable assumptions on matrix sparsity, input access, and the condition number of $A$~\cite{hhl}.
Subsequent developments have further improved the precision dependence and resource requirements of quantum linear-system algorithms~\cite{cks, costa, qls_review}. 
Nevertheless, the circuit cost of such coherent algorithms remains a significant challenge for large-scale implementation on noisy quantum processors.

Variational quantum algorithms have also been explored as an alternative approach to solving linear systems~\cite{cqs, vqls, xu_vqls, poisson_vqa, dynamic_vqls}.
One approach reformulates the linear-system problem as a variational ground-state problem for a cost Hamiltonian.
For a normalized state $|b\rangle$ and an invertible matrix $A$, the cost Hamiltonian $H_b(A)=A^\dagger(I-|b\rangle\langle b|)A$ has the normalized solution $|x\rangle=A^{-1}|b\rangle/\|A^{-1}|b\rangle\|$ as its unique zero-energy ground state, up to a global phase~\cite{cost_hamiltonian, xu_vqls}. 
This formulation provides a variational cost for assessing the quality of approximate solutions and enables iterative refinement through optimization. 
However, the scalability of variational quantum algorithms can be limited by measurement costs, barren plateaus, and sensitivity to noise~\cite{bp1,bp2,bp3,mea_cost}.

The cost Hamiltonian formulation also allows other quantum eigensolvers to be applied to linear-system problems. 
Several recent quantum eigensolvers sample computational-basis states on quantum hardware and classically diagonalize the Hamiltonian projected onto the resulting subspace~\cite{qsci, qsci_hci, sqd}.
Among these approaches, sample-based quantum diagonalization (SQD) has demonstrated promising results in quantum chemistry by combining quantum measurements with classical diagonalization, symmetry-based configuration recovery, and iterative subspace refinement~\cite{sqd, lsqd, half_mobius, protein_ligand,afqmc, embedding, piegen, csqd, extsqd, sqd_photochemistry}.
In terms of sampling strategies, sample-based Krylov quantum diagonalization (SKQD) uses quantum Krylov states for computational-basis sampling~\cite{skqd}, while SqDRIFT extends this approach through randomized Hamiltonian simulation~\cite{sqdrift, qdrift}.
These developments motivate applying SQD-type subspace diagonalization to the cost Hamiltonian $H_b(A)$.

However, applying quantum Krylov-state sampling to linear-system problems presents several challenges. 
Even when $A$ admits an efficiently implementable Hamiltonian decomposition, directly expanding $A^\dagger A$ can increase the number of terms quadratically, while $A^\dagger|b\rangle\langle b|A$ may introduce additional quantum implementation overhead. 
Consequently, generating quantum Krylov states through direct simulation of $H_b(A)$ can be costly in quantum resources. 
An alternative is to generate quantum Krylov states $e^{-iAT}|b\rangle$ using the original Hermitian operator $A$.
Previous quantum linear-system algorithms have used linear combinations of such time-evolution operators to approximate the action of $A^{-1}$ on $|b\rangle$~\cite{cks,cqs,xu_krylov}.
However, these approaches do not directly establish a computational-basis sampling protocol with guarantees for identifying the dominant components of the solution from measurements of individual Krylov states.

In this work, we propose a sample-based quantum linear solver that combines quantum Krylov-state sampling with classical variational diagonalization. 
The proposed method measures quantum Krylov states generated by time evolution under the original Hermitian operator $A$ in the computational basis to collect candidate basis states.
The cost Hamiltonian $H_b(A)$ is then projected onto the selected subspace and diagonalized classically to reconstruct an approximate solution.

To justify this sampling procedure, we exploit a quantum Krylov-state representation of the inverse operator to derive lower bounds on the sampling probabilities of computational-basis states in terms of their amplitudes in the solution. 
Based on this connection, we establish support-discovery guarantees for two sampling schemes: finite nonnegative-time sampling selects the evolution time $T$ from a finite nonnegative grid, whereas signed-time sampling allows both positive and negative evolution times.
The former imposes an upper bound on the evolution time, whereas the latter achieves improved ideal shot complexity at the cost of occasionally requiring long time evolutions.
Specifically, for a condition-number bound $\kappa$ on $A$, a lower bound $\beta_L$ on the solution probability weight of each of $L$ target basis states, and a failure probability $\delta$, signed-time sampling achieves a shot complexity of $O(\kappa^2\beta_L^{-1}\log(L/\delta))$. 
We show that this scaling matches a worst-case lower bound up to constant factors within the evolution-only observation model in a specified parameter regime.
In addition, we analyze the effects of finite-length qDRIFT compilation~\cite{qdrift} and circuit reuse on sampling and derive an infidelity bound for the reconstructed solution.

Although approximate solutions can also be reconstructed through restricted least-squares optimization within the selected computational-basis subspace, we adopt projected diagonalization to maintain a direct connection with the variational framework of SQD. 
In particular, the variational energy of $H_b(A)$, together with a lower bound on the singular values of $A$, provides an upper bound on the solution infidelity, and projected diagonalization minimizes this bound within the selected subspace.
The variational cost also serves as a common objective for evaluating and comparing different candidate subspaces. 
These properties make the formulation suitable for iterative refinement through SQD-type basis-state selection and carryover, as well as recovery procedures, particularly when the dimension of the classical diagonalization subspace is constrained.

The remainder of this paper is organized as follows. Section~\ref{sec:algorithm} presents the proposed sample-based quantum linear solver, including quantum Krylov-state sampling and classical variational reconstruction. Section~\ref{sec:theory} establishes the main theoretical results concerning coordinate discovery, randomized compilation, retained solution weight, variational reconstruction, and sampling lower bounds. Detailed proofs and additional analyses are provided in the appendices.

\section{Algorithm overview}\label{sec:algorithm}
\subsection{Problem setup and variational reconstruction}\label{sec:problem}
Let $A=A^\dagger\in\mathbb C^{N\times N}$ be invertible, with $N=2^n$ and $\|b\|=1$.
Non-Hermitian linear systems can be reduced to the Hermitian case through the standard Hermitian dilation, which preserves the condition number~\cite{hhl}.
We seek the normalized solution
\begin{equation}
 |x\rangle=\frac{A^{-1}|b\rangle}{\|A^{-1}b\|}.
 \label{eq:target}
\end{equation}
The normalized solution $|x\rangle$ is the unique zero-energy ground state, up to a global phase, of the cost Hamiltonian~\cite{cost_hamiltonian, xu_vqls}
\begin{equation}
 H_b(A)=A^\dagger(I-|b\rangle\langle b|)A.
 \label{eq:cost-hamiltonian}
\end{equation}

We assume the ability to prepare $|b\rangle$ and implement time evolution under each term in a decomposition of $A$ of the form
\begin{equation}
 \begin{gathered}
 A=\sum_{j=1}^{m}a_jh_j,\qquad
 \lambda_A=\sum_{j=1}^{m}|a_j|,\\
 a_j\in\mathbb R,\qquad h_j=h_j^\dagger,\qquad \|h_j\|\leq1.
 \end{gathered}
 \label{eq:decomposition}
\end{equation}
For selected computational basis strings $z,z'$, classical reconstruction requires evaluating $\langle Az|Az'\rangle$ and $\langle Az|b\rangle$.
A sufficient condition is efficiently enumerable sparse columns of $A$, together with classical amplitude queries to $b$.

For a nonempty set of selected strings $\mathcal S$, let $E_{\mathcal S}$ have the corresponding computational-basis vectors as columns, and put $P_{\mathcal S}=E_{\mathcal S}E_{\mathcal S}^\dagger$.
For classical reconstruction, we form
\begin{equation}
 H_{\mathcal S}=E_{\mathcal S}^\dagger H_b(A)E_{\mathcal S}
 \label{eq:projected-cost}
\end{equation}
and use a normalized lowest eigenvector $u$ of $H_{\mathcal S}$ to obtain
\begin{equation}
 |\widehat x\rangle=E_{\mathcal S}u,\qquad
 e_{\mathcal S}=u^\dagger H_{\mathcal S}u.
 \label{eq:ritz-output}
\end{equation}
The output is a sparse classical representation of a normalized approximate solution state.
The diagonalization minimizes the same variational cost for every chosen subspace.
The resulting energy provides a criterion for comparing candidate subspaces and, with a spectral lower bound, an upper bound on solution infidelity, as detailed in Sec.~\ref{sec:reconstruction}.

A practical implementation may impose an upper limit on $|\mathcal S|$ based on available classical resources.
The resulting cost can guide iterative subspace selection, potentially incorporating native symmetry-based or code-space recovery~\cite{sqd,csr}, multiple sampling batches, and carryover of useful basis states.

\subsection{Quantum Krylov-state sampling}\label{sec:fresh}\label{sec:reuse-protocol}

Quantum Krylov-state sampling collects candidate basis strings by preparing $|b\rangle$, evolving it under $A$ for a selected time $T$ to obtain $e^{-iAT}|b\rangle$ under ideal evolution, and measuring the resulting state in the computational basis.
The collected basis strings provide candidates for the subspace used in classical variational diagonalization.

We consider two schemes for selecting the evolution time $T$.
In finite nonnegative-time sampling, an integer index $k\in\{0,\ldots,d-1\}$ is selected with probability $\pi_k$, and the evolution time is set to $T=kt$ for $t>0$.
The theoretical guarantee in Sec.~\ref{sec:finite} uses the uniform distribution $\pi_k=1/d$.
In signed-time sampling, $T\in\mathbb R$ is drawn from the probability density $p_a(T)$ defined in Sec.~\ref{sec:signed}, allowing both positive and negative evolution times.

Time evolution $e^{-iAT}$ can optionally be approximated using qDRIFT~\cite{qdrift}.
When qDRIFT is used, states are prepared using randomized circuit realizations and measured individually.
A qDRIFT realization is determined by the selected evolution time $T$, a positive integer $D$, and randomly sampled term indices.
The corresponding circuit is given by
\begin{equation}
 V_\omega(T)=
 e^{-i\lambda_AT\operatorname{sgn}(a_{j_D})h_{j_D}/D}
 \cdots
 e^{-i\lambda_AT\operatorname{sgn}(a_{j_1})h_{j_1}/D},
 \label{eq:qdrift}
\end{equation}
where $\omega=(j_1,\ldots,j_D)$ consists of independent draws with probabilities $|a_j|/\lambda_A$.
The integer $D$ denotes the number of sampled terms in a single qDRIFT circuit and may vary with the selected time $T$.
Negative times reverse the signs of the term rotations, and $T=0$ requires no evolution.

When qDRIFT is used, we distinguish fresh-randomization sampling from circuit reuse according to how circuit realizations are generated and measured.
In fresh-randomization sampling, each shot begins by selecting an evolution time $T$.
For finite nonnegative-time sampling, $k$ is selected with probability $\pi_k$ and $T=kt$, whereas signed-time sampling draws $T$ directly from its probability density.
A new qDRIFT realization $V_\omega(T)$ is then generated for the selected time.
We prepare $|b\rangle$, apply the circuit, and measure once in the computational basis to obtain a single basis string.
This procedure is repeated independently for $M$ shots to collect basis strings.

In contrast, circuit reuse generates $R_k$ independent qDRIFT realizations at each predetermined evolution time $T_k$.
The number of realizations $R_k$ and the number of executions per realization $s_k$ are chosen according to available computational resources and the desired sampling accuracy, and are fixed before circuit generation.
For each realization, the preparation of $|b\rangle$, circuit application, and computational-basis measurement are repeated $s_k$ times.
The total number of measurement shots and independent circuit draws are, respectively,
\begin{equation}
 M=\sum_kR_ks_k,\qquad \sum_kR_k.
 \label{eq:reuse-budget}
\end{equation}
For each of the $d$ evolution times, generating $R$ independent qDRIFT realizations and executing each realization $s$ times gives $M=dRs$ measurement shots from $dR$ independent circuit draws.

\section{Theoretical guarantees}\label{sec:theory}
We assume supplied spectral bounds
\begin{equation}
 \begin{gathered}
 \operatorname{spec}(A)\subseteq[-B,-a]\cup[a,B],\\
 0<a\leq B,\qquad \kappa=B/a.
 \end{gathered}
 \label{eq:spectrum}
\end{equation}
The bounds may be conservative, and $\kappa$ need not equal the exact condition number.
In particular, $B=\lambda_A$ is a valid choice under Eq.~\eqref{eq:decomposition}.
Unless stated otherwise, we assume exact state preparation and term rotations.
Asymptotic shot and total-work bounds are stated for $0<\delta\leq1/2$, where $\delta$ denotes the target failure probability.
The explicit bounds, including integer ceilings, remain valid for $0<\delta<1$.

To analyze support discovery, let $x_z=\langle z|x\rangle$ denote the computational-basis amplitude of the normalized solution state $|x\rangle$ at basis string $z$.
For a chosen $1\leq L\leq N$, order the coordinates by decreasing $|x_z|$, breaking ties deterministically, and let $\mathcal T_L$ contain the first $L$.
For complete-support discovery, assume
\begin{equation}
 \sum_{z\in\mathcal T_L}|x_z|^2\geq\alpha_L,
 \qquad |x_z|^2\geq\beta_L>0\quad(z\in\mathcal T_L).
 \label{eq:concentration}
\end{equation}
Thus $\alpha_L$ is a lower bound on retained solution weight, and $\beta_L$ is a lower bound on the squared amplitude $|x_z|^2$ of each target coordinate, with $L\beta_L\leq1$.
This concentration convention follows the setting in Refs.~\cite{skqd,sqdrift}.
Let $\mathcal S_M$ denote the set of distinct computational-basis strings observed in $M$ measurement shots.
The following support-discovery bounds establish conditions under which $\Pr(\mathcal T_L\subseteq\mathcal S_M)\geq1-\delta$, while the reconstruction bounds depend on the solution weight actually retained in the selected subspace.

\subsection{General sampling guarantees}\label{sec:transfer}
A linear combination of quantum Krylov states can approximate the solution, but our algorithm does not prepare this combination.
We establish conditions under which measurements of the individual states suffice to discover the target basis strings.

Suppose an analytical vector $v$ and a probability distribution $q$ over computational-basis strings satisfy
\begin{equation}
 \|v-x\|\leq\epsilon,\qquad |v_z|^2\leq C^2q(z)\quad\text{for every }z,
 \label{eq:envelope}
\end{equation}
with $C>0$ and $\epsilon\geq0$. 
The first condition controls approximation of the solution.
The second relates the coordinate amplitudes of $v$ to the actual measurement law through a bounded coefficient factor.

\begin{proposition}[General sampling bounds]\label{prop:transfer}
Under Eq.~\eqref{eq:envelope}, every $z\in\mathcal T_L$ satisfies
\begin{equation}
 q(z)\geq\frac{[\sqrt{\beta_L}-\epsilon]_+^2}{C^2},
 \label{eq:sampling-floor}
\end{equation}
where $[u]_+=\max\{u,0\}$.
The coordinate probability bound is nontrivial when $\epsilon<\sqrt{\beta_L}$.
For $M$ independent shots from $q$,
\begin{equation}
 \Pr(\mathcal T_L\subseteq\mathcal S_M)
 \geq1-L\exp\!\left(-\frac{M[\sqrt{\beta_L}-\epsilon]_+^2}{C^2}\right).
 \label{eq:discovery}
\end{equation}
For any set $\mathcal R$ of computational-basis strings, let $P_{\mathcal R}$ denote the projector onto their span.
Equation~\eqref{eq:envelope} also implies 
\begin{equation}
 \|P_{\mathcal R}x\|\leq\epsilon+C\sqrt{q(\mathcal R)},
 \qquad q(\mathcal R)=\sum_{z\in\mathcal R}q(z).
 \label{eq:set-envelope}
\end{equation}
\end{proposition}

For a finite time mixture with coefficients not all zero, take
\begin{equation}
 \begin{split}
 v&=\sum_kc_ke^{-iAT_k}b,\\
 q_\pi(z)&=\sum_k\pi_k|\langle z|e^{-iAT_k}|b\rangle|^2,\\
 C_\pi^2&=\sum_{k:\pi_k>0}\frac{|c_k|^2}{\pi_k},
 \end{split}
 \label{eq:finite-witness}
\end{equation}
where $\pi_k>0$ whenever $c_k\neq0$. Weighted Cauchy--Schwarz supplies the coordinate inequality in Eq.~\eqref{eq:envelope}. Uniform sampling gives $C_\pi^2=d\sum_k|c_k|^2$; for fixed coefficients, choosing $\pi_k$ proportional to $|c_k|$ minimizes this factor.

Krylov and inverse-approximation methods construct accurate representations of $A^{-1}b$ in suitable spans~\cite{saad,cks}. Sampling requires more: a representation with controlled coefficient size. The following constructions provide that control without resolving individual eigenvalues. The proof of Proposition~\ref{prop:transfer} is given in Appendix~\ref{app:transfer}.

\subsection{Support discovery with finite nonnegative times}\label{sec:finite}
A uniform mixture over a finite list of evolution times is a simple sampling rule. Its spacing must avoid phase aliasing, and its length must make the inverse witness accurate without excessively increasing its coefficient factor. The next result leaves these choices free within explicit sufficient conditions.

\begin{theorem}[Finite-time sampling]\label{thm:finite}
Assume Eq.~\eqref{eq:spectrum}. Choose $0<t\leq\pi/B$ and positive integers $J,\ell$ such that
\begin{equation}
 J\sin(at/2)>1,\qquad d=J\ell,
 \label{eq:finite-admissibility}
\end{equation}
and define the filter-error bound
\begin{equation}
 \epsilon_{\mathrm f}
 =Bt+(1+Bt)[J\sin(at/2)]^{-\ell}.
 \label{eq:finite-parameters}
\end{equation}
Choose $k$ uniformly from $0,\ldots,d-1$ and measure $e^{-ikAt}|b\rangle$. The outcome law obeys
\begin{equation}
 q(z)\geq\frac{[|x_z|-\epsilon_{\mathrm f}]_+^2}{(Bdt)^2}
 \quad\text{for every }z.
 \label{eq:finite-floor}
\end{equation}
The largest evolution time is $(d-1)t$. For $\pi_k=1/d$, the finite representation in Eq.~\eqref{eq:finite-witness} admits coefficients satisfying
\begin{equation}
 \begin{gathered}
 \left\|\sum_kc_ke^{-ikAt}b-x\right\|\leq\epsilon_{\mathrm f},\\
 \sum_k|c_k|\leq Bdt,\qquad C_\pi\leq Bdt.
 \end{gathered}
 \label{eq:finite-coefficients}
\end{equation}
\end{theorem}

Here $J$ is the averaging-window length used to construct the witness, and $\ell$ determines how often that averaging is repeated in the proof. The processor only samples uniformly from the resulting $d$ times. 
The condition $Bt\leq\pi$ keeps the spectral phases in an interval where their distance from zero controls the geometric sum. The quantity $[J\sin(at/2)]^{-1}<1$ bounds its contraction.

Whenever $\epsilon_{\mathrm f}<\sqrt{\beta_L}$, Proposition~\ref{prop:transfer} gives the explicit sufficient budget
\begin{equation}
 M\geq\left\lceil
 \frac{(Bdt)^2}{(\sqrt{\beta_L}-\epsilon_{\mathrm f})^2}
 \ln\frac L\delta\right\rceil
 \label{eq:finite-shots}
\end{equation}
for complete discovery with probability at least $1-\delta$. 
This bound displays the trade-off between a longer time list, which improves the filter, and its coefficient factor.

There are admissible choices with
\begin{equation}
 \begin{gathered}
 t=\Theta(\sqrt{\beta_L}/B),\qquad J=\Theta((at)^{-1}),\\
 \ell=\Theta(1+\ln(1/\beta_L)),
 \end{gathered}
 \label{eq:finite-support-choice}
\end{equation}
whose filter error is at most a fixed fraction of $\sqrt{\beta_L}$. These give
\begin{equation}
 \begin{gathered}
 M=O\!\left(\frac{\kappa^2\ell^2}{\beta_L}\ln\frac L\delta\right),\\
 d=O(\kappa\ell/\sqrt{\beta_L}),\qquad
 (d-1)t=O(\ell/a).
 \end{gathered}
 \label{eq:finite-scaling}
\end{equation}
Appendix~\ref{app:finite} provides constructive choices and proves these orders. Neither the target coordinates nor $\|A^{-1}b\|$ are needed to execute them, although the complete-support budget uses a supplied lower bound on $\beta_L$.

The proof constructs a damped geometric series in $e^{-iAt}$. Averaging consecutive powers contracts the residual uniformly over the allowed spectral intervals, and repeated convolution reduces that residual without producing large coefficients. This simultaneously controls approximation error and the measurement-probability bound. The construction depends on separation from zero, not gaps between distinct eigenvalues; positive definiteness and a nonzero overlap $\langle x|b\rangle$ are unnecessary.

\subsection{Signed-time sampling and ideal shot complexity}\label{sec:signed}
Allowing both signs of the evolution time gives a different trade-off. Rather than discretize a one-sided inverse approximation, we use an absolutely integrable representation of $A^{-1}$ over real times~\cite{hastingswen,bachmann}. Define $\operatorname{sinc}(u)=\sin(u)/u$, with its continuous value at zero, and
\begin{align}
 h_a(T)&=\frac{10a}{33\pi}\operatorname{sinc}^6(aT/6),\nonumber\\
 s_a(T)&=\operatorname{sgn}(T)\int_{|T|}^{\infty}h_a(u)\,du,\nonumber\\
 \Lambda_a&=\int_{\mathbb R}|s_a(T)|\,dT,\qquad
 p_a(T)=\frac{|s_a(T)|}{\Lambda_a}.
 \label{eq:signed-density}
\end{align}
The value of $s_a$ at zero is immaterial. The density $p_a$ depends only on $a$, not on the solution. Its normalization and moments are evaluated in Appendix~\ref{app:signed}. They satisfy $\Lambda_a=\Theta(a^{-1})$ and $\mathbb E_{p_a}T^2=\Theta(a^{-2})$, with universal constants determined by the specified filter.

\begin{theorem}[Signed-time coordinate domination]\label{thm:signed}
Assume Eq.~\eqref{eq:spectrum}. A shot that draws $T\sim p_a$, prepares $|b\rangle$, and measures after exact evolution has outcome law
\begin{equation}
 q_a(z)=\int_{\mathbb R}p_a(T)|\langle z|e^{-iAT}|b\rangle|^2\,dT.
 \label{eq:signed-law}
\end{equation}
It satisfies
\begin{equation}
 |x_z|^2\leq
 \frac{\Lambda_a^2}{\|A^{-1}b\|^2}q_a(z)
 \leq(B\Lambda_a)^2q_a(z)
 \quad\text{for all }z.
 \label{eq:signed-domination}
\end{equation}
Consequently, for $0<\delta<1$,
\begin{equation}
 M\geq\left\lceil\frac{(B\Lambda_a)^2}{\beta_L}\ln\frac L\delta\right\rceil
 \label{eq:signed-shots}
\end{equation}
independent shots recover $\mathcal T_L$ with probability at least $1-\delta$.
\end{theorem}

The identity behind this bound is
\begin{equation}
 A^{-1}=i\int_{\mathbb R}s_a(T)e^{-iAT}\,dT.
 \label{eq:exact-inverse}
\end{equation}
The Fourier transform of $h_a$ vanishes for $|\omega|\geq a$, so the scalar integral equals $1/\lambda$ on the entire allowed spectrum. Taking a coordinate of the corresponding vector identity and applying Cauchy--Schwarz proves Eq.~\eqref{eq:signed-domination}. 
Appendix~\ref{app:time-sampler} gives an explicit rejection sampler for $p_a$ with a constant expected number of proposals. Bounded-time truncation and time rounding are treated in Appendix~\ref{app:time-cutoff}.

Compared with the finite-grid scaling in Eq.~\eqref{eq:finite-scaling}, the ideal shot count has no squared logarithmic factor. The basic signed-time distribution has unbounded support, however. Its finite second moment controls expected compilation cost, not a deterministic maximum depth. A truncated version retains the shot scaling up to constants while introducing a different worst-case time and depth budget. Section~\ref{sec:lower} states the precise observation-model lower bound matched by Eq.~\eqref{eq:signed-shots}.

\subsection{Randomized compilation and circuit reuse}\label{sec:compilation}
\subsubsection{Fresh circuits}
The qDRIFT channel and the accuracy of a single qDRIFT realization are different objects~\cite{qdrift,chen}. For the product in Eq.~\eqref{eq:qdrift}, independence of its terms gives the first-moment estimate
\begin{equation}
 \|\mathbb E_\omega V_\omega(T)-e^{-iAT}\|
 \leq\frac{\lambda_A^2T^2}{D}.
 \label{eq:first-moment}
\end{equation}
The bound holds for signed times and requires no small-step assumption. 
A direct proof, together with second-moment and channel estimates, is given in Appendix~\ref{app:compilation}.

For an ideal finite witness in Eq.~\eqref{eq:finite-witness}, replace each evolution by a random unitary $V_k$ and bound its mean-compilation error by
\begin{equation}
 \sum_k|c_k|\|(\mathbb E V_k-e^{-iAT_k})b\|
 \leq\epsilon_{\mathrm{av}}.
 \label{eq:average-error}
\end{equation}
The actual average outcome law is
$\widetilde q_\pi(z)=\sum_k\pi_k\mathbb E|\langle z|V_k|b\rangle|^2$.
If the ideal witness error is at most $\epsilon$, then
\begin{equation}
 \widetilde q_\pi(z)\geq
 \frac{[|x_z|-\epsilon-\epsilon_{\mathrm{av}}]_+^2}{C_\pi^2}.
 \label{eq:compiled-transfer}
\end{equation}
Indeed, the mean witness approximates $x$ with error at most $\epsilon+\epsilon_{\mathrm{av}}$, and Cauchy--Schwarz applies on the joint time--circuit probability space. The integral version follows by the same argument.

For the schedule of Theorem~\ref{thm:finite}, let $T_{\max}=(d-1)t$. A uniform positive integer length $D$ for the nonzero times gives
\begin{equation}
 \widetilde q(z)\geq
 \frac{[|x_z|-\epsilon_{\mathrm f}
       -Bdt\lambda_A^2T_{\max}^2/D]_+^2}{(Bdt)^2},
 \label{eq:finite-compiled-floor}
\end{equation}
where $\widetilde q$ is the uniform-time averaged outcome law. For any allocated mean-compilation error $\epsilon_{\mathrm{av}}>0$, a sufficient uniform length is
\begin{equation}
 D\geq\max\!\left\{1,
 \left\lceil\frac{Bdt\,\lambda_A^2T_{\max}^2}
                        {\epsilon_{\mathrm{av}}}\right\rceil\right\}.
 \label{eq:finite-depth}
\end{equation}
If $\epsilon_{\mathrm f}+\epsilon_{\mathrm{av}}<\sqrt{\beta_L}$, then fresh compiled shots recover the targets with probability at least $1-\delta$ whenever
\begin{equation}
 M\geq\left\lceil
 \frac{(Bdt)^2}{(\sqrt{\beta_L}-\epsilon_{\mathrm f}-\epsilon_{\mathrm{av}})^2}
 \ln\frac L\delta\right\rceil.
 \label{eq:finite-compiled-shots}
\end{equation}
Thus the filter and compilation errors share the available amplitude margin.

For the signed-time sampler, choose any witness-error budget $\epsilon>0$ and use
\begin{equation}
 D(T)=\max\!\left\{1,
 \left\lceil\frac{B\Lambda_a\lambda_A^2T^2}{\epsilon}\right\rceil\right\}.
 \label{eq:signed-depth}
\end{equation}
The compiled law satisfies Eq.~\eqref{eq:envelope} with error at most $\epsilon$ and $C=B\Lambda_a$. For $0<\epsilon<\sqrt{\beta_L}$, a sufficient complete-discovery budget is
\begin{equation}
 M\geq\left\lceil
 \frac{(B\Lambda_a)^2}{(\sqrt{\beta_L}-\epsilon)^2}
 \ln\frac L\delta\right\rceil.
 \label{eq:signed-compiled-shots}
\end{equation}
The expected term count per shot obeys
\begin{equation}
 \mathbb E_{p_a}D(T)\leq1+
 \frac{B\Lambda_a\lambda_A^2}{\epsilon}\,
 \mathbb E_{p_a}T^2.
 \label{eq:expected-depth}
\end{equation}
Appendix~\ref{app:compilation} proves these budgets and the corresponding total term counts.

To compare resource orders, choose the finite-grid parameters as in Eq.~\eqref{eq:finite-support-choice} and keep the total witness error below a fixed fraction of $\sqrt{\beta_L}$. 
The finite-time sufficient length then scales as
$O((\lambda_A/a)^2\kappa\ell^3/\sqrt{\beta_L})$.
For signed times, allocating $\epsilon$ as a fixed fraction of $\sqrt{\beta_L}$ gives expected length
$O(1+(\lambda_A/a)^2\kappa/\sqrt{\beta_L})$.
Requiring a uniform channel-error estimate to preserve a fixed fraction of the finite-grid coordinate floor instead gives the larger sufficient length
$O((\lambda_A/a)^2\kappa^2\ell^4/\beta_L)$.
The constructive choices and total term counts are given in Appendix~\ref{app:compilation}.

Additional physical noise can be treated through its effect on the measurement law. If the actual independent-shot law has total-variation distance at most $\eta$ from a certified law with target floor $q_*$, complete discovery has probability at least $1-L e^{-M(q_*-\eta)}$ when $\eta<q_*$. A trace-distance bound on the corresponding averaged output states supplies such a total-variation bound. A more forgiving retained-weight version is given in Appendix~\ref{app:mass}; it introduces an accuracy floor rather than requiring preservation of every small coordinate probability.

\subsubsection{Circuit pools}
Repeated measurements of a fixed circuit share its randomness. For its coordinate probability $p_\omega(z)$, the miss probability is
$\mathbb E(1-p_\omega(z))^s$, not $(1-\mathbb E p_\omega(z))^s$.
To analyze the protocol in Sec.~\ref{sec:reuse-protocol}, we control the representativeness of the full circuit pool rather than require every circuit to approximate the ideal state.

\begin{theorem}[Stratified circuit reuse]\label{thm:reuse}
At fixed times $T_k$, let $V_k$ be random unitaries and suppose deterministic coefficients $c_k$, not all zero, satisfy
\begin{equation}
 \left\|\sum_kc_k\mathbb E V_kb-x\right\|\leq\epsilon.
 \label{eq:reuse-witness}
\end{equation}
Fix positive integers $R_k,s_k$ before drawing the pool. Draw $R_k$ independent realizations at time $k$ and measure each $s_k$ times. All circuit draws are mutually independent, and all measurements are conditionally independent given the circuits. Set
\begin{equation}
 M=\sum_kR_ks_k,\qquad
 \pi_k=\frac{R_ks_k}{M},\qquad
 C_\pi^2=\sum_k\frac{|c_k|^2}{\pi_k}.
 \label{eq:reuse-weights}
\end{equation}
For $0<\zeta<1$, define
\begin{equation}
 \Delta_R=
 \left(\sum_k\frac{|c_k|^2}{R_k}\right)^{1/2}
 \left(1+\sqrt{2\ln(1/\zeta)}\right).
 \label{eq:pool-error}
\end{equation}
Then
\begin{equation}
 \begin{split}
 &\Pr(\mathcal T_L\subseteq\mathcal S_M)\\
 &\quad\geq1-\zeta-
 L\exp\!\left(-\frac{M[\sqrt{\beta_L}-\epsilon-\Delta_R]_+^2}
                       {C_\pi^2}\right).
 \end{split}
 \label{eq:reuse-support}
\end{equation}
\end{theorem}

The two sampling resources enter differently. For equal allocations, the squared prefactor inside Eq.~\eqref{eq:pool-error} is $C_\pi^2/(dR)$, whereas the observation budget is $M=dRs$. 
Increasing shots per circuit improves coverage of the given pool but does not reduce its representation error. 
The proof uses a bounded-differences estimate~\cite{mcdiarmid} for an empirical witness and then conditions on the pool. It is provided in Appendix~\ref{app:compilation}, together with a random-time pool version and alternative sufficient conditions.

\subsection{Retained solution weight and variational reconstruction}\label{sec:reconstruction}
Complete support discovery is useful for an exactly sparse solution, but stronger than necessary for approximation. Small coordinates can remain unobserved without contributing much error. Define the actual best-$L$ tail
\begin{equation}
 \tau_L=\sum_{z\notin\mathcal T_L}|x_z|^2.
 \label{eq:tail}
\end{equation}
It obeys $\tau_L\leq1-\alpha_L$ when Eq.~\eqref{eq:concentration} holds.

\begin{theorem}[Retained-weight guarantee]\label{thm:mass}
Suppose Eq.~\eqref{eq:envelope} holds and $\mathcal S_M$ contains all outcomes of $M\geq1$ independent shots from $q$. For a fixed $1\leq L\leq N$ and $0<\delta<1$, with probability at least $1-\delta$,
\begin{equation}
 \begin{split}
 &1-\|P_{\mathcal S_M}x\|^2\\
 &\quad\leq\min\!\left\{1,\tau_L+
 \left(\epsilon+C\sqrt{\frac{L\ln2+\ln(1/\delta)}{M}}\right)^2
 \right\}.
 \end{split}
 \label{eq:mass-bound}
\end{equation}
No minimum nonzero coordinate weight is required. A simultaneous statement for every $L$ holds after replacing the numerator by $L\ln2+\ln[L(L+1)/\delta]$ and minimizing the right-hand side over $L$.
\end{theorem}

The proof bounds the probability that an unobserved subset of $\mathcal T_L$ has appreciable sampling mass, then uses Eq.~\eqref{eq:set-envelope} to convert sampling mass to solution mass. 
This is a finite-set occupancy argument, related to the general study of missing mass~\cite{benhamou}; the complete proof and an expectation version appear in Appendix~\ref{app:mass}. For circuit reuse, the same bound holds with $C=C_\pi$ and $\epsilon$ replaced by $\epsilon+\Delta_R$, with total failure probability at most $\delta+\zeta$. Conditional miss probabilities, rather than unconditional independent-shot probabilities, justify this extension.

The weight captured by a coordinate set controls its variational solution independently of how the set was obtained.

\begin{proposition}[Variational reconstruction and certificate]\label{prop:ritz}
Under Eq.~\eqref{eq:spectrum},
\begin{equation}
 a^2(I-|x\rangle\langle x|)\preceq H_b(A)
 \preceq B^2(I-|x\rangle\langle x|).
 \label{eq:sandwich}
\end{equation}
For any nonempty $\mathcal S$, an exact normalized lowest Ritz vector $\widehat x$ from Eq.~\eqref{eq:ritz-output} satisfies
\begin{align}
 \langle\widehat x|H_b(A)|\widehat x\rangle
 &\leq B^2(1-\|P_{\mathcal S}x\|^2),\label{eq:ritz-energy}\\
 1-|\langle x|\widehat x\rangle|^2
 &\leq\min\{1,\kappa^2(1-\|P_{\mathcal S}x\|^2)\}.
 \label{eq:ritz-fidelity}
\end{align}
For every normalized candidate $y$, not necessarily a Ritz vector,
\begin{equation}
 1-|\langle x|y\rangle|^2
 \leq\min\!\left\{1,\frac{\langle y|H_b(A)|y\rangle}{a^2}\right\}.
 \label{eq:certificate}
\end{equation}
\end{proposition}

The cost certificate is the VQLS-type residual-to-state relation~\cite{vqls,cost_hamiltonian}; Eq.~\eqref{eq:ritz-fidelity} additionally connects it to retained coordinate weight. The nonzero spectrum of $H_b(A)$ lies in $[a^2,B^2]$. Testing its Rayleigh quotient on the normalized truncation of $x$ proves the variational upper bound. Appendix~\ref{app:mass} includes the proof, including the case $P_{\mathcal S}x=0$.
For the same selected subspace and exact classical solves, the Ritz energy certificate is no larger than the basic condition-number residual certificate of least-squares reconstruction; Appendix~\ref{app:least-squares} proves this comparison.

If the reconstruction subspace contains $\mathcal T_L$, Eq.~\eqref{eq:ritz-fidelity} gives infidelity at most $\min\{1,\kappa^2(1-\alpha_L)\}$. Alternatively, Theorem~\ref{thm:mass} controls the reconstruction on the full observation subspace $\mathcal S_M$ without $\beta_L$. For a target infidelity $0<\eta<1$, suppose
\begin{equation}
 \tau_L<\frac{\eta}{\kappa^2},\qquad
 0\leq\epsilon<\sqrt{\eta/\kappa^2-\tau_L}.
 \label{eq:accuracy-margin}
\end{equation}
Then the explicit budget
\begin{equation}
 M\geq\left\lceil
 \frac{C^2[L\ln2+\ln(1/\delta)]}
 {\bigl(\sqrt{\eta/\kappa^2-\tau_L}-\epsilon\bigr)^2}
 \right\rceil
 \label{eq:accuracy-budget}
\end{equation}
ensures that exact Ritz reconstruction on $\mathcal S_M$ has infidelity at most $\eta$ with probability at least $1-\delta$. A supplied upper bound on $\tau_L$ can be substituted in these conditions. The remaining accuracy margin, after accounting for the tail and witness error, determines the sufficient sampling budget. Selecting a smaller batch from $\mathcal S_M$ does not automatically preserve its retained-weight guarantee; Eq.~\eqref{eq:certificate} still applies to the candidate actually obtained.

When $\kappa^2\tau_L$ stays below $\eta$ by a fixed fractional margin, the signed-time construction with $C=B\Lambda_a$ allows a witness budget of order $\sqrt\eta/\kappa$ within Eq.~\eqref{eq:accuracy-margin}. The resulting sufficient scales are
\begin{equation}
 \begin{split}
 M&=O\!\left(\frac{\kappa^4}{\eta}[L+\ln(1/\delta)]\right),\\
 \mathbb E D&=O\!\left(1+
      \frac{\kappa^2}{\sqrt\eta}\left(\frac{\lambda_A}{a}\right)^2\right).
 \end{split}
 \label{eq:accuracy-scaling}
\end{equation}
Here the chosen witness error must leave a fixed fraction of the remaining margin for sampling. Appendix~\ref{app:mass} gives explicit sufficient allocations and the finite-grid counterpart.

Inexact classical reconstruction can be handled separately. If the computed Hermitian matrix differs from $H_{\mathcal S}$ by operator norm at most $\xi_H$, and the returned normalized vector has Rayleigh quotient at most $\xi_{\mathrm{eig}}$ above its lowest eigenvalue, then
\begin{equation}
 1-|\langle x|\widehat x\rangle|^2
 \leq\min\!\left\{1,
 \kappa^2(1-\|P_{\mathcal S}x\|^2)
 +\frac{2\xi_H+\xi_{\mathrm{eig}}}{a^2}\right\}.
 \label{eq:finite-precision}
\end{equation}
A reliably evaluated cost also certifies the final candidate through Eq.~\eqref{eq:certificate}, regardless of its proposal mechanism. 
Certified selection and pruning procedures are recorded in Appendix~\ref{app:selection}.

The classical cost depends on the size of the selected coordinate set. If each selected column of $A$ has at most $s_A$ efficiently enumerable nonzeros, their row union has at most $s_A|\mathcal S|$ entries. After the columns are constructed and sorted, their pairwise inner products form $H_{\mathcal S}$ in $O(s_A|\mathcal S|^2)$ arithmetic operations, with at most $s_A|\mathcal S|$ amplitude queries to $b$. Dense diagonalization costs $O(|\mathcal S|^3)$ arithmetic operations and $O(|\mathcal S|^2)$ matrix storage. The full columns must be retained in this calculation: projecting $A$ and then squaring it generally gives the wrong projected cost. Appendix~\ref{app:mass} supplies the exact formula, a compact singular-value implementation, and the relation to sampled-column least squares.

\subsection{Sampling lower bounds and limitations}\label{sec:lower}
Concentration alone does not make an arbitrary time grid useful. For $A=X$ and $b=|0\rangle$, the solution is $|1\rangle$ and $\kappa=1$. At every integer multiple of $t=\pi$,
$e^{-ikAt}|b\rangle=(-1)^k|0\rangle$.
The target is never observed, even with exact evolution. The single-term qDRIFT circuit has the same obstruction. 
The constructive choices in Appendix~\ref{app:finite-choice} ensure a positive target-amplitude margin, while the signed-time law avoids relying on a single resonant grid.

There is also an intrinsic sampling cost even when arbitrary evolution times are allowed. The following statements concern protocols whose samples are computational-basis measurements of fresh $e^{-iAT}|b\rangle$ states and whose output support contains only observed coordinates. The time may depend on previous outcomes.

\begin{theorem}[Evolution-sampling lower bounds]\label{thm:lower}
Let $L\geq2$ and $\kappa\geq1$.
For every $0<\beta\leq1/[2(L-1)]$ and $0<\delta\leq1/4$, there is an invertible Hermitian system with an exactly $L$-supported solution and minimum target weight $\beta$ for which complete support discovery with probability at least $1-\delta$ requires
\begin{equation}
 M=\Omega\!\left(\frac{\kappa^2}{\beta}\ln\frac L\delta\right).
 \label{eq:support-lower}
\end{equation}
For every $0<\eta\leq1/8$ and $0<\delta\leq1/4$, there is an exactly $L$-supported solution for which
$\Pr(1-\|P_{\mathcal S_M}x\|^2\leq\eta)\geq1-\delta$
requires
\begin{equation}
 M=\Omega\!\left(\frac{\kappa^2}{\eta}[L+\ln(1/\delta)]\right).
 \label{eq:mass-lower}
\end{equation}
Both statements allow adaptive signed times and remain valid for qDRIFT using a diagonal decomposition of the constructed instance.
\end{theorem}

The instances are diagonal. Evolution changes their phases but not their measurement probabilities, and conditioning suppresses the probabilities of important solution coordinates by a factor of order $\kappa^2$. Recovering all rare coordinates incurs a coupon-collection logarithm; retaining most of their total weight gives Eq.~\eqref{eq:mass-lower}. Appendix~\ref{app:lower} proves both claims with explicit constants.

Equation~\eqref{eq:signed-shots} matches Eq.~\eqref{eq:support-lower} up to universal constants in the stated parameter range. For an exactly supported solution and exact signed-time evolution, Theorem~\ref{thm:mass} also matches the retained-mass scaling in Eq.~\eqref{eq:mass-lower}. This does not close the gap to the sufficient final-Ritz budget in Eq.~\eqref{eq:accuracy-scaling}.

\section{Discussion}\label{sec:discussion}

We have proposed a sample-based quantum linear solver that combines quantum Krylov-state sampling with classical variational diagonalization.
Rather than preparing the solution state or implementing time evolution under the cost Hamiltonian $H_b(A)$, the method measures quantum Krylov states generated by the original operator $A$ to collect candidate basis strings.
The projected cost Hamiltonian is then diagonalized classically to reconstruct an approximate solution.
This division of tasks assigns the discovery of candidate basis states to the quantum processor and coefficient determination to the classical processor.
Our support-discovery bounds establish conditions under which important basis strings can be observed without preparing the solution state.
The reconstruction bounds relate the solution weight retained in the selected subspace to the accuracy of the reconstructed state.
Together, these results provide separate guarantees for the sampling and reconstruction stages.

The two Krylov-state sampling schemes involve different resource trade-offs.
Finite nonnegative-time sampling bounds the maximum evolution time, making the longest required evolution predictable.
Signed-time sampling allows both positive and negative evolution times and can achieve improved ideal shot complexity, but may occasionally require long evolutions.
When qDRIFT is used, approximation errors due to finite circuit length and the cost of generating circuit realizations must also be considered.
Circuit reuse can reduce the number of independently generated circuit realizations by executing each realization multiple times, but the number of realizations and the number of measurement shots play distinct statistical roles.

The applicability of the method depends on the structure of the solution state and the information available for classical reconstruction.
In particular, solution concentration can depend on the choice of basis, and structural properties of $A$, such as sparsity and locality, do not guarantee that the solution amplitudes are concentrated.
Even when candidate basis strings can be sampled efficiently, classical input access and the costs of constructing and diagonalizing the projected cost Hamiltonian should also be considered.
Future work should identify problem classes where the proposed method could offer a quantum advantage and develop sampling strategies tailored to their structure.

\begin{acknowledgments}
This work was supported by the National Research Foundation of Korea (NRF) grants funded by the Korean government (MSIT) (nos. RS-2025-00519428 and RS-2026-25612745) and by an Institute of Information \& Communications Technology Planning \& Evaluation (IITP) grant funded by the Korean government (MSIT) (no. RS-2025-25464788).
D.A. also acknowledges the National Quantum Laboratory (QLab) at the University of Maryland and the Cleveland Clinic Quantum Innovation Catalyzer Program for providing supportive research environments.
ChatGPT was used to assist with mathematical derivations and manuscript writing; the authors take full responsibility for the final content.

\end{acknowledgments}

\section*{Author contributions}
B. P. conceived the idea for this work, developed the algorithm, derived the theoretical results, and wrote the manuscript. K. J. and D. A. supervised the work.

\section*{Data availability}
All analytical results and their derivations are presented in this article and its appendices.

\section*{Competing interests}
D.A. is affiliated with Singularity Quantum Inc., and the other authors declare no competing interests.

\appendix

\section{Sampling-transfer arguments}\label{app:transfer}
\subsection{Proof of Proposition~\ref{prop:transfer}}
For a target coordinate $z\in\mathcal T_L$, the norm error controls its amplitude error:
\begin{equation}
 |v_z|\geq |x_z|-|v_z-x_z|\geq\sqrt{\beta_L}-\epsilon.
\end{equation}
Combining this inequality with $|v_z|^2\leq C^2q(z)$ proves Eq.~\eqref{eq:sampling-floor}, including the case of a nonpositive lower amplitude bound. Independence implies that a fixed target is unobserved with probability $(1-q(z))^M\leq e^{-Mq(z)}$. Summing over the $L$ targets gives Eq.~\eqref{eq:discovery}. For a general coordinate set $\mathcal R$,
\begin{equation}
 \|P_{\mathcal R}x\|\leq\|P_{\mathcal R}(x-v)\|+
 \|P_{\mathcal R}v\|
 \leq\epsilon+C\sqrt{q(\mathcal R)},
\end{equation}
which proves Eq.~\eqref{eq:set-envelope}.

\subsection{Finite and continuous witnesses}
For the finite representation in Eq.~\eqref{eq:finite-witness}, weighted Cauchy--Schwarz gives
\begin{equation}
 \begin{split}
 |v_z|^2
 &=\left|\sum_{k:\pi_k>0}\frac{c_k}{\sqrt{\pi_k}}
             \sqrt{\pi_k}\langle z|e^{-iAT_k}|b\rangle\right|^2\\
 &\leq\left(\sum_k\frac{|c_k|^2}{\pi_k}\right)
          \sum_k\pi_k|\langle z|e^{-iAT_k}|b\rangle|^2.
 \end{split}
 \label{eq:weighted-cs}
\end{equation}
A zero-probability term has zero coefficient and is omitted. Another application of Cauchy--Schwarz shows
$(\sum_k|c_k|)^2\leq\sum_k|c_k|^2/\pi_k$,
with equality for $\pi_k=|c_k|/\sum_j|c_j|$ on the nonzero coefficients. This establishes the coefficient-optimal sampling rule for a fixed representation.

The same reasoning applies to a normalized ensemble $\psi_\omega$ with probability measure $\mu$. If
$v=\int f(\omega)\psi_\omega\,d\mu(\omega)$ and
$\int|f|^2\,d\mu\leq C^2$, then
\begin{equation}
 |v_z|^2\leq C^2\int|\langle z|\psi_\omega\rangle|^2\,d\mu(\omega).
 \label{eq:ensemble-cs}
\end{equation}
The integral defining $v$ exists because the states are normalized and $f$ is integrable by Cauchy--Schwarz. This formulation includes continuous time laws and joint time--circuit randomness.

\section{The finite nonnegative-time construction}\label{app:finite}
\subsection{Proof of Theorem~\ref{thm:finite}}
The proof uses a polynomial in $U=e^{-iAt}$ to approximate the inverse action on $b$. Its coefficients must remain controlled, because they will also bound the coordinate probabilities. Start with the averaging operator
\begin{equation}
 F=\frac1J\sum_{r=1}^JU^r.
 \label{eq:averaging-polynomial}
\end{equation}
For every eigenvalue $\lambda$ of $A$, we have
$0<at\leq|\lambda|t\leq Bt\leq\pi$.
On this interval $\sin(|\lambda|t/2)\geq\sin(at/2)>0$. The geometric-sum identity therefore gives
\begin{equation}
 \left|\frac1J\sum_{r=1}^Je^{-ir\lambda t}\right|
 =\frac{|\sin(J\lambda t/2)|}{J|\sin(\lambda t/2)|}
 \leq\frac1{J\sin(at/2)}<1.
 \label{eq:F-contraction}
\end{equation}
The spectral theorem bounds $\|F^\ell\|$ by $[J\sin(at/2)]^{-\ell}$. The same phase condition shows that $I-U$ is invertible.

To damp a geometric series without increasing its coefficients, let $R_1,\ldots,R_\ell$ be independent uniform integers in $\{1,\ldots,J\}$. Within this proof, write $W=R_1+\cdots+R_\ell$ and $g_k=\Pr(W>k)$ for $0\leq k<d=J\ell$. 
Since $W\leq d$,
\begin{equation}
 \begin{split}
 \sum_{k=0}^{d-1}g_kU^k
 &=\mathbb E\sum_{k=0}^{W-1}U^k\\
 &=(I-U)^{-1}(I-\mathbb E U^W)\\
 &=(I-U)^{-1}(I-F^\ell).
 \end{split}
 \label{eq:finite-series-identity}
\end{equation}
The first equality follows by expanding the indicators $\mathbf1_{\{W>k\}}$ in a finite sum. The last equality uses independence and the fact that all factors are powers of the same operator. Thus averaging the stopping point of the geometric series gives a residual $F^\ell$, while every coefficient $g_k$ stays in $[0,1]$. Although $F^\ell$ contains $U^d$, the left side uses only $U^0,\ldots,U^{d-1}$.

Set $\nu=\|A^{-1}b\|$ locally, so $b=\nu Ax$ and $\nu^{-1}\leq B$. Define
\begin{equation}
 c_k=\frac{it}{\nu}g_k,\qquad
 v=\sum_{k=0}^{d-1}c_kU^kb.
 \label{eq:finite-proof-witness}
\end{equation}
All factors in Eq.~\eqref{eq:finite-series-identity} commute with $A$, so that identity implies
\begin{equation}
 v=itA(I-e^{-iAt})^{-1}(I-F^\ell)x.
 \label{eq:finite-proof-identity}
\end{equation}
There are two approximation errors: the factor $itA(I-e^{-iAt})^{-1}$ need not be the identity, and $F^\ell$ is the residual of the damped series. For the first error, the integral Taylor remainder gives
$|e^{-i\lambda t}-1+i\lambda t|\leq(\lambda t)^2/2$.
Since $|\sin(\lambda t/2)|\geq|\lambda|t/\pi$ in the allowed interval,
\begin{equation}
 \left|\frac{i\lambda t}{1-e^{-i\lambda t}}-1\right|
 \leq\frac{(\lambda t)^2}{4|\sin(\lambda t/2)|}
 \leq\frac{\pi|\lambda|t}{4}\leq|\lambda|t.
 \label{eq:finite-scalar-error}
\end{equation}
It follows that $itA(I-e^{-iAt})^{-1}$ differs from the identity by at most $Bt$ and has norm at most $1+Bt$. Subtracting $x$ in Eq.~\eqref{eq:finite-proof-identity} and using Eq.~\eqref{eq:F-contraction} yields
\begin{equation}
 \|v-x\|\leq Bt+(1+Bt)[J\sin(at/2)]^{-\ell}
 =\epsilon_{\mathrm f}.
 \label{eq:finite-proof-error}
\end{equation}
Finally, $0\leq g_k\leq1$ and $\nu^{-1}\leq B$ imply
\begin{equation}
 \sum_k|c_k|\leq Bdt,\qquad
 d\sum_k|c_k|^2\leq(Bdt)^2.
 \label{eq:finite-proof-coefficients}
\end{equation}
Apply Eq.~\eqref{eq:weighted-cs} with $\pi_k=1/d$ and then the coordinate triangle inequality to obtain Eq.~\eqref{eq:finite-floor}. The last time in the list is $(d-1)t$, as stated.

\subsection{Constructive accuracy and resource choices}\label{app:finite-choice}
For any desired filter error $0<\epsilon\leq1$, choose $0<Bt<\epsilon$ and any integer $J$ satisfying $J\sin(at/2)>1$. A sufficient integer repetition count is
\begin{equation}
 \ell\geq\max\!\left\{1,
 \left\lceil
 \frac{\ln[(1+Bt)/(\epsilon-Bt)]}{\ln[J\sin(at/2)]}
 \right\rceil\right\}.
 \label{eq:finite-error-choice}
\end{equation}
Its denominator is positive by the contraction condition. 
Substitution in Eq.~\eqref{eq:finite-parameters} proves $\epsilon_{\mathrm f}\leq\epsilon$ without fixing a particular error split.

To establish the resource orders, one possible choice is
$t=\epsilon/(2B)$ and $J=\lceil2/\sin(at/2)\rceil$, followed by Eq.~\eqref{eq:finite-error-choice} with its smallest allowed $\ell$. It gives $J\sin(at/2)\geq2$. The bounds $\sin(at/2)\geq at/\pi$ and $\sin(at/2)\leq at/2$ imply
\begin{equation}
 \begin{gathered}
 J=\Theta(\kappa/\epsilon),\qquad
 \ell=\Theta(1+\ln(1/\epsilon)),\\
 d=O(\kappa\ell/\epsilon),\qquad
 dt=O(\ell/a),\qquad Bdt=O(\kappa\ell).
 \end{gathered}
 \label{eq:finite-accuracy-orders}
\end{equation}
For the claimed order of $\ell$, also note that $J\sin(at/2)<2+\sin(at/2)\leq3$, so the denominator in Eq.~\eqref{eq:finite-error-choice} is bounded above and below by positive universal constants.

Taking, for example, $\epsilon=\sqrt{\beta_L}/2$ leaves an amplitude margin of at least $\sqrt{\beta_L}/2$ in Eq.~\eqref{eq:finite-shots}. Equation~\eqref{eq:finite-accuracy-orders} then proves Eqs.~\eqref{eq:finite-support-choice} and \eqref{eq:finite-scaling}. The integer ceiling is absorbed into the shot order because $\ln(L/\delta)\geq\ln2$ for the stated asymptotic range $\delta\leq1/2$.

The witness also permits a nonuniform discrete schedule proportional to $g_k$, since
$\sum_kg_k=\mathbb EW=\ell(J+1)/2$.
The unknown normalization $\nu$ cancels from these probabilities. The uniform schedule in the main text avoids computing the survival weights.

\section{Signed-time inverse sampling}\label{app:signed}
\subsection{Normalization, moments, and the inverse identity}
We use the Fourier convention
$\widehat f(\omega)=\int_{\mathbb R}f(T)e^{-i\omega T}\,dT$.
Let $Z=\int_{\mathbb R}\operatorname{sinc}^6u\,du$ for this appendix. The identity
$\operatorname{sinc}(u)=\tfrac12\int_{-1}^1e^{iuv}\,dv$
expresses $\operatorname{sinc}^6$ as the characteristic function of a sum of six independent uniform variables on $[-1,1]$. The density of that sum at zero is
\begin{equation}
 \frac{6^5-6\,4^5+15\,2^5}{2^6\,5!}=\frac{11}{40}.
\end{equation}
This value follows by inclusion--exclusion for the volume of a sliced cube, or directly by sixfold convolution. Fourier inversion gives $Z=11\pi/20$. The same convolution has support $[-6,6]$ and is continuous and zero at its endpoints. Thus
\begin{equation}
 h_a(T)=\frac{a}{6Z}\operatorname{sinc}^6(aT/6)
\end{equation}
is an even probability density with $\widehat h_a(\omega)=0$ for $|\omega|\geq a$, including the endpoints.

The necessary absolute moments can be evaluated explicitly. Using
\begin{equation}
 \sin^6u=\frac{10-15\cos(2u)+6\cos(4u)-\cos(6u)}{32},
\end{equation}
four integrations by parts give
\begin{equation}
 \begin{split}
 &\int_{\mathbb R}|u|\operatorname{sinc}^6u\,du\\
 &\quad=-\frac58\int_0^\infty\frac{\cos(2u)-\cos(4u)}u\,du\\
 &\qquad+\frac{27}{8}\int_0^\infty\frac{\cos(4u)-\cos(6u)}u\,du\\
 &\quad=\frac{27}{8}\ln3-4\ln2.
 \end{split}
 \label{eq:signed-moment-one}
\end{equation}
Two integrations by parts similarly yield
\begin{equation}
 \begin{split}
 &\int_{\mathbb R}|u|^3\operatorname{sinc}^6u\,du\\
 &\quad=\frac{15}{8}\int_0^\infty\frac{\cos(2u)-\cos(4u)}u\,du\\
 &\qquad-\frac98\int_0^\infty\frac{\cos(4u)-\cos(6u)}u\,du\\
 &\quad=3\ln2-\frac98\ln3.
 \end{split}
 \label{eq:signed-moment-three}
\end{equation}
The boundary terms vanish at zero because $\sin^6u=O(u^6)$, and at infinity because the differentiated numerators are bounded. The final integrals are evaluated as differences using
$\int_0^\infty[\cos(ru)-\cos(su)]\,du/u=\ln(s/r)$ for $r,s>0$.

Tonelli's theorem, followed by these moments, gives
\begin{align}
 \Lambda_a
 &=\int_{\mathbb R}|u|h_a(u)\,du
 =\frac{6}{Za}\left(\frac{27}{8}\ln3-4\ln2\right)
,\label{eq:Lambda-exact}\\
 \mathbb E_{p_a}T^2
 &=\frac{\int_{\mathbb R}|u|^3h_a(u)\,du}{3\Lambda_a}
 =\frac{12(3\ln2-\tfrac98\ln3)}
       {a^2(\tfrac{27}{8}\ln3-4\ln2)}
.
 \label{eq:signed-second-moment}
\end{align}
Both integrals are finite since $\operatorname{sinc}^6u\leq\min\{1,|u|^{-6}\}$.

For every $\lambda\neq0$, oddness of $s_a$ and Fubini's theorem imply
\begin{equation}
 \begin{split}
 i\int_{\mathbb R}s_a(T)e^{-i\lambda T}\,dT
 &=2\int_0^\infty h_a(u)\int_0^u\sin(\lambda T)\,dT\,du\\
 &=\frac{1-\widehat h_a(\lambda)}{\lambda}.
 \end{split}
 \label{eq:signed-scalar-inverse}
\end{equation}
Fubini is justified by $\int|u|h_a(u)\,du<\infty$. On the spectrum in Eq.~\eqref{eq:spectrum}, the right-hand side is $1/\lambda$. Applying the finite-dimensional spectral theorem proves Eq.~\eqref{eq:exact-inverse}; its operator integral converges absolutely because $s_a\in L^1$ and $\|e^{-iAT}\|=1$.

\subsection{Proof of Theorem~\ref{thm:signed}}
The inverse identity supplies the exact vector representation
\begin{equation}
 x=\frac{\Lambda_a}{\|A^{-1}b\|}
       \mathbb E_{T\sim p_a}[i\operatorname{sgn}(T)e^{-iAT}b].
 \label{eq:signed-vector-identity}
\end{equation}
Equation~\eqref{eq:ensemble-cs} bounds every coordinate by
$(\Lambda_a/\|A^{-1}b\|)^2q_a(z)$.
Since $\|A^{-1}b\|\geq1/B$, Eq.~\eqref{eq:signed-domination} follows. Apply Proposition~\ref{prop:transfer} with $\epsilon=0$ and $C=B\Lambda_a$ to obtain Eq.~\eqref{eq:signed-shots}. Equation~\eqref{eq:Lambda-exact} gives $B\Lambda_a=\Theta(\kappa)$ and hence the asserted ideal shot scaling.

\subsection{Sampling the time distribution}\label{app:time-sampler}
The density $p_a$ can be sampled without evaluating the integral in $s_a$. Draw a proposal $v>0$ as $\sqrt U$ with probability $2/3$, and as $U^{-1/4}$ otherwise, where $U$ is uniform on $(0,1)$. Accept with probability $\operatorname{sinc}^6v$ when $v\leq1$, and $\sin^6v$ when $v>1$. After acceptance, draw $T$ uniformly on $[-6v/a,6v/a]$.

The positive proposal has density
$\tfrac43v$ on $(0,1)$ and $\tfrac43v^{-5}$ on $(1,\infty)$. Multiplication by its acceptance probability gives
$\tfrac43v\operatorname{sinc}^6v$ on the whole positive axis. Its total acceptance probability is
\begin{equation}
 \frac23\left(\frac{27}{8}\ln3-4\ln2\right).
\end{equation}
Upon setting $r=6v/a$, the accepted radius has density $2rh_a(r)/\Lambda_a$. A uniform conditional time in $[-r,r]$ consequently has density
\begin{equation}
 \int_{|T|}^{\infty}\frac{2rh_a(r)}{\Lambda_a}\frac{dr}{2r}
 =\frac{|s_a(T)|}{\Lambda_a}=p_a(T).
\end{equation}
All draws in this procedure are independent when repeated for fresh shots.

\subsection{Bounded-time and finite-precision variants}\label{app:time-cutoff}
A finite time cutoff can be chosen from its actual omitted coefficient mass. For $T_c>0$, write
\begin{equation}
 \Lambda_{a,c}=\int_{|T|\leq T_c}|s_a(T)|\,dT.
 \label{eq:retained-coefficient-mass}
\end{equation}
The elementary bound on $\operatorname{sinc}$ gives
\begin{equation}
 \begin{split}
 \Lambda_a-\Lambda_{a,c}
 &=2\int_{T_c}^{\infty}(u-T_c)h_a(u)\,du\\
 &\leq\frac{6^5}{10Za^5T_c^4},
 \end{split}
 \label{eq:signed-tail}
\end{equation}
where $Z=11\pi/20$ as above. For a total witness budget $0<\epsilon<1$, choose a cutoff satisfying
\begin{equation}
 B(\Lambda_a-\Lambda_{a,c})<\epsilon
 \label{eq:time-cutoff}
\end{equation}
and sample $p_a$ conditioned on $|T|\leq T_c$. The truncated ideal witness uses coefficient mass $\Lambda_{a,c}/\|A^{-1}b\|$ and differs from $x$ by at most $B(\Lambda_a-\Lambda_{a,c})$. This leaves the remaining positive margin for compilation. Set
\begin{equation}
 D_c(T)=\max\!\left\{1,
 \left\lceil
 \frac{B\Lambda_{a,c}\lambda_A^2T^2}
 {\epsilon-B(\Lambda_a-\Lambda_{a,c})}
 \right\rceil\right\}.
 \label{eq:cutoff-depth}
\end{equation}
Equation~\eqref{eq:first-moment}, integrated against the truncated witness, bounds the additional error by the denominator in Eq.~\eqref{eq:cutoff-depth}. Consequently the compiled law satisfies Eq.~\eqref{eq:envelope} with total error at most $\epsilon$ and coefficient factor $C=B\Lambda_{a,c}\leq B\Lambda_a$.

Because the conditional density is $|s_a(T)|/\Lambda_{a,c}$ on the retained interval, its normalization cancels from the expected-cost estimate:
\begin{equation}
 \mathbb E[D_c(T)\mid |T|\leq T_c]\leq1+
 \frac{B\lambda_A^2\int_{|T|\leq T_c}T^2|s_a(T)|\,dT}
 {\epsilon-B(\Lambda_a-\Lambda_{a,c})}.
 \label{eq:cutoff-expected}
\end{equation}
The maximum length has the same denominator and numerator $B\Lambda_{a,c}\lambda_A^2T_c^2$, plus one.

For resource orders, it is enough to make the omitted witness error at most a fixed fraction of $\epsilon$. For instance, the right-hand side of Eq.~\eqref{eq:signed-tail} can be set to $\epsilon/(2B)$. This gives a cutoff of order $a^{-1}(\kappa/\epsilon)^{1/4}$ and implies
\begin{equation}
 \begin{split}
 \mathbb E[D_c(T)\mid |T|\leq T_c]
 &=O\!\left(1+\left(\frac{\lambda_A}{a}\right)^2
                         \frac{\kappa}{\epsilon}\right),\\
 \max_{|T|\leq T_c}D_c(T)
 &=O\!\left(1+\left(\frac{\lambda_A}{a}\right)^2
                    \kappa^{3/2}\epsilon^{-3/2}\right).
 \end{split}
 \label{eq:cutoff-maximum}
\end{equation}
The probability that a draw from $p_a$ passes this cutoff is $\Lambda_{a,c}/\Lambda_a\geq1-\epsilon/(2B\Lambda_a)>1/2$. Here $B\Lambda_a\geq1$ follows from the inverse identity and $\|A^{-1}\|\geq1/B$. Thus rejection at the cutoff adds only a constant expected number of proposals for this construction. Taking $\epsilon$ as a fixed fraction of $\sqrt{\beta_L}$ preserves the compiled support-shot scaling, with a different maximum-depth dependence from the finite grid.

If a retained time is rounded to $\widetilde T$ with
$|T-\widetilde T|\leq\Delta/2$, then
$\|e^{-iAT}-e^{-iA\widetilde T}\|\leq B\Delta/2$.
Keeping the analytical coefficient associated with the original time, the extra witness error is at most $B^2\Lambda_{a,c}\Delta/2$. Reserve this amount from the total budget and assign compilation lengths to the rounded times. For the untruncated law, replace $\Lambda_{a,c}$ by $\Lambda_a$. An additional perturbation of the time distribution can instead be handled through the induced outcome law as in Appendix~\ref{app:mass}.

\section{Randomized compilation and circuit-pool proofs}\label{app:compilation}
\subsection{qDRIFT moment estimates}
For a fixed real time $T$, put $\tau=T/D$ and
\begin{equation}
 \overline U_\tau=\sum_j\frac{|a_j|}{\lambda_A}
       e^{-i\lambda_A\tau\operatorname{sgn}(a_j)h_j},
 \qquad U_\tau=e^{-iA\tau}.
\end{equation}
For a Hermitian matrix $G$ and real $s$, the integral remainder formula and unitarity give
\begin{equation}
 \|e^{-isG}-I+isG\|\leq\frac{s^2\|G\|^2}{2}.
 \label{eq:taylor-matrix}
\end{equation}
The linear terms of $\overline U_\tau$ and $U_\tau$ coincide. Since $\|A\|\leq\lambda_A$,
$\|\overline U_\tau-U_\tau\|\leq\lambda_A^2\tau^2$.
Both matrices are contractions. Independence of the qDRIFT terms implies $\mathbb E V_\omega(T)=\overline U_\tau^D$, and a telescoping sum proves
\begin{equation}
 \|\overline U_\tau^D-U_\tau^D\|\leq D\|\overline U_\tau-U_\tau\|
 \leq\frac{\lambda_A^2T^2}{D}.
\end{equation}
This is Eq.~\eqref{eq:first-moment}, a standard mean-product estimate~\cite{chen}.

Unitarity also gives the state second-moment bound
\begin{equation}
 \begin{split}
 \mathbb E\|(V_\omega(T)-e^{-iAT})b\|^2
 &=2-2\operatorname{Re}\langle b|e^{iAT}\mathbb E V_\omega(T)|b\rangle\\
 &\leq\frac{2\lambda_A^2T^2}{D}.
 \end{split}
 \label{eq:qdrift-second}
\end{equation}
For a density matrix $\rho$, define the averaged circuit channel by
$\mathcal E_{T,D}(\rho)=\mathbb E V_\omega(T)\rho V_\omega(T)^\dagger$.
The generator of unitary conjugation by $G$ is
$\mathcal L_G(\rho)=-i[G,\rho]$ and has induced trace norm at most $2\|G\|$. Unitary conjugation preserves the trace norm. Applying an integral remainder to these channels bounds each second-order remainder by $2s^2\|G\|^2$. The exact and averaged one-step channels have the same first-order term, so their difference has induced norm at most $4\lambda_A^2\tau^2$. Telescoping the channels yields
\begin{equation}
 \frac12\|\mathcal E_{T,D}(\rho)-e^{-iAT}\rho e^{iAT}\|_1
 \leq\frac{2\lambda_A^2T^2}{D}.
 \label{eq:qdrift-channel}
\end{equation}
The argument remains valid with an ancillary system. These moment and channel estimates are consistent with the qDRIFT and random-product-formula analyses~\cite{qdrift,chen}.

\subsection{Averaged-amplitude transfer and resource budgets}
Let $v=\sum_kc_ke^{-iAT_k}b$ be an ideal witness, and put
$\overline v=\sum_kc_k\mathbb E V_kb$.
The triangle inequality gives
$\|\overline v-x\|\leq\epsilon+\epsilon_{\mathrm{av}}$.
On the joint distribution that draws $k$ with probability $\pi_k$ and then draws its circuit,
\begin{equation}
 \begin{split}
 |\overline v_z|^2
 &=\left|\sum_k\pi_k\mathbb E\left[
              \frac{c_k}{\pi_k}\langle z|V_k|b\rangle\right]\right|^2\\
 &\leq C_\pi^2\sum_k\pi_k\mathbb E|\langle z|V_k|b\rangle|^2.
 \end{split}
\end{equation} 
The coordinate triangle inequality proves Eq.~\eqref{eq:compiled-transfer}.

For the finite construction, Eqs.~\eqref{eq:first-moment} and \eqref{eq:finite-coefficients} give
\begin{equation}
 \sum_k|c_k|\|(\mathbb EV_k-e^{-iAT_k})b\|
 \leq Bdt\,\frac{\lambda_A^2T_{\max}^2}{D}.
\end{equation}
The length in Eq.~\eqref{eq:finite-depth} makes this no larger than the allocated $\epsilon_{\mathrm{av}}$. Substituting the displayed error bound into Eq.~\eqref{eq:compiled-transfer} gives Eq.~\eqref{eq:finite-compiled-floor}. Adding the filter error and applying Proposition~\ref{prop:transfer} proves Eq.~\eqref{eq:finite-compiled-shots}.

For signed times, use the mean witness
\begin{equation}
 \overline v=\frac{\Lambda_a}{\|A^{-1}b\|}
       \mathbb E_{T,\omega}[i\operatorname{sgn}(T)V_\omega(T)b].
\end{equation}
Its coefficient magnitude is at most $B\Lambda_a$. Equation~\eqref{eq:signed-depth} bounds the mean-evolution error at each time by $\epsilon/(B\Lambda_a)$, including the zero-error case $T=0$. Integrating proves $\|\overline v-x\|\leq\epsilon$. Cauchy--Schwarz on the joint time--circuit measure then gives
$|\overline v_z|^2\leq(B\Lambda_a)^2\widetilde q_a(z)$.
Proposition~\ref{prop:transfer} establishes Eq.~\eqref{eq:signed-compiled-shots}. Finally,
$\max\{1,\lceil u\rceil\}\leq1+u$ for $u\geq0$ proves Eq.~\eqref{eq:expected-depth}. At $T=0$ the implementation may omit the evolution altogether.

The resource orders can be established with explicit positive error margins. In the finite construction of Appendix~\ref{app:finite-choice}, set the filter budget and the mean-compilation budget both to $\sqrt{\beta_L}/4$. Their sum is at most $\sqrt{\beta_L}/2$, while $Bdt=O(\kappa\ell)$ and $T_{\max}=O(\ell/a)$, with $\ell=\Theta(1+\ln(1/\beta_L))$. Equations~\eqref{eq:finite-depth} and \eqref{eq:finite-compiled-shots} yield
\begin{equation}
 D=O\!\left(\left(\frac{\lambda_A}{a}\right)^2
                   \frac{\kappa\ell^3}{\sqrt{\beta_L}}\right)
\end{equation}
and total term count
\begin{equation}
 MD=O\!\left(\left(\frac{\lambda_A}{a}\right)^2
             \frac{\kappa^3\ell^5}{\beta_L^{3/2}}\ln\frac L\delta\right).
 \label{eq:finite-total-cost}
\end{equation}

For the signed-time construction, take $\epsilon=\sqrt{\beta_L}/2$. The exact identities in Eqs.~\eqref{eq:Lambda-exact} and \eqref{eq:signed-second-moment} give $B\Lambda_a=\Theta(\kappa)$ and $\mathbb E_{p_a}T^2=\Theta(a^{-2})$. Therefore the total expected term count is
\begin{equation}
 M\mathbb ED(T)=
 O\!\left(\frac{\kappa^2}{\beta_L}\ln\frac L\delta
       \left[1+\left(\frac{\lambda_A}{a}\right)^2
                     \frac{\kappa}{\sqrt{\beta_L}}\right]\right).
 \label{eq:signed-total-cost}
\end{equation}
These are counts of the specified term exponentials, not elementary gates.

For comparison, the exact finite-grid law has target floor
$q_*=(\sqrt{\beta_L}-\epsilon_{\mathrm f})^2/(Bdt)^2$.
To preserve a fixed positive fraction of this floor by directly subtracting the channel error in Eq.~\eqref{eq:qdrift-channel}, its upper bound must be at most a fixed fraction of $q_*$. This gives the sufficient order
\begin{equation}
 D=O(\lambda_A^2T_{\max}^2/q_*)
 =O\!\left(\left(\frac{\lambda_A}{a}\right)^2
              \frac{\kappa^2\ell^4}{\beta_L}\right)
\end{equation}
for the above finite-grid choices.

\subsection{Proof of Theorem~\ref{thm:reuse}}
Write $\psi_{k,r}=V_{k,r}b$ and define
\begin{equation}
 v_R=\sum_k\frac{c_k}{R_k}\sum_{r=1}^{R_k}\psi_{k,r},
 \qquad \overline v=\sum_kc_k\mathbb E\psi_{k,r}.
\end{equation}
The centered summands are independent. Their cross terms vanish in the expected squared Hilbert-space norm, and $\|\psi_{k,r}\|=1$, so
\begin{equation}
 \mathbb E\|v_R-\overline v\|^2\leq\sum_k\frac{|c_k|^2}{R_k},
 \qquad
 \mathbb E\|v_R-\overline v\|\leq
       \left(\sum_k\frac{|c_k|^2}{R_k}\right)^{1/2}.
\end{equation}
Changing one circuit at time $k$ changes the real-valued function
$\|v_R-\overline v\|$ by at most $2|c_k|/R_k$.
McDiarmid's bounded-differences inequality~\cite{mcdiarmid} therefore implies
\begin{equation}
 \begin{split}
 &\Pr\!\left(\|v_R-\overline v\|>
       \sqrt{\sum_k|c_k|^2/R_k}+u\right)\\
 &\quad\leq\exp\!\left(-\frac{u^2}{2\sum_k|c_k|^2/R_k}\right).
 \end{split}
\end{equation}
The coefficients are not all zero, so the denominator is positive. Choosing $u$ from failure probability $\zeta$ gives
$\|v_R-x\|\leq\epsilon+\Delta_R$ with probability at least $1-\zeta$.

Condition on a pool satisfying that event and put
\begin{equation}
 q_R(z)=\sum_k\frac{\pi_k}{R_k}
             \sum_{r=1}^{R_k}|\langle z|\psi_{k,r}\rangle|^2.
\end{equation}
Weighted Cauchy--Schwarz gives $|(v_R)_z|^2\leq C_\pi^2q_R(z)$. For any fixed coordinate set $\mathcal J$, the conditional probability that all measurements miss it is
\begin{equation}
 \begin{split}
 \prod_{k,r}(1-p_{k,r}(\mathcal J))^{s_k}
 &\leq\exp\!\left(-\sum_{k,r}s_kp_{k,r}(\mathcal J)\right)\\
 &=e^{-Mq_R(\mathcal J)},
 \end{split}
 \label{eq:conditional-miss}
\end{equation}
where $p_{k,r}(\mathcal J)=\|P_{\mathcal J}\psi_{k,r}\|^2$. The last equality uses the allocation $\pi_k=R_ks_k/M$, not an independent-shot assumption. Apply the coordinate argument of Proposition~\ref{prop:transfer}, sum over targets, and add the pool failure probability $\zeta$. This proves Eq.~\eqref{eq:reuse-support}. Equation~\eqref{eq:conditional-miss} also supplies the conditional input to the retained-mass proof below.

\subsection{Random-time pools and alternative reuse criteria}
For a pool whose circuits include the random choice of time, suppose $C_\infty>0$, a normalized output $\psi_\omega$, and a deterministic function $f$ of the circuit randomness satisfy
\begin{equation}
 |f(\omega)|\leq C_\infty,\qquad
 \|\mathbb E[f(\omega)\psi_\omega]-x\|\leq\epsilon.
\end{equation}
Draw $R$ independent circuits and measure each $s$ times. The empirical witness
$R^{-1}\sum_rf(\omega_r)\psi_{\omega_r}$ has mean squared deviation at most $C_\infty^2/R$, and replacing one draw changes its norm deviation by at most $2C_\infty/R$. The preceding proof therefore gives a pool error
\begin{equation}
 \frac{C_\infty}{\sqrt R}\left(1+\sqrt{2\ln(1/\zeta)}\right).
 \label{eq:random-pool-error}
\end{equation}
The support and retained-weight formulas use this error, $C=C_\infty$, and $M=Rs$. For signed-time sampling one may take
$f(T,\omega)=i\operatorname{sgn}(T)\Lambda_a/\|A^{-1}b\|$
and $C_\infty=B\Lambda_a$. For a finite mixture, $f(k,\omega)=c_k/\pi_k$ works when its magnitudes have a uniform bound. The unknown normalization is needed only for the witness, not for circuit generation.

An alternative sufficient condition starts from individual circuit accuracy. Suppose an ideal finite witness has error $\epsilon<\sqrt{\beta_L}$ and coefficient mass $\Gamma=\sum_k|c_k|$. For each target, at least one time satisfies
$|\langle z|e^{-iAT_k}|b\rangle|\geq(\sqrt{\beta_L}-\epsilon)/\Gamma$.
If each circuit at each time differs from its ideal output by norm at most $\epsilon_Q$ with probability at least $1-\zeta$, let
\begin{equation}
 p=\left[\frac{\sqrt{\beta_L}-\epsilon}{\Gamma}-\epsilon_Q\right]_+^2.
\end{equation}
With $R$ independent circuits at every time and $s$ measurements per circuit, a fixed target is missed by its favorable time with probability at most
$[\zeta+(1-\zeta)(1-p)^s]^R$.
Summing over targets gives success probability at least
$1-L[\zeta+(1-\zeta)(1-p)^s]^R$.
Equation~\eqref{eq:qdrift-second} and Markov's inequality supply the sufficient uniform length
$D\geq2\lambda_A^2T_{\max}^2/(\zeta\epsilon_Q^2)$.

A coordinate-specific second moment gives another option. For a circuit ensemble let
$\mu_z=\mathbb Ep_\omega(z)$ and
$\chi_z=\mathbb Ep_\omega(z)^2/\mu_z^2$ for $\mu_z>0$.
If $\mu_z\geq q_*>0$ and $\chi_z\leq\chi$ on the targets, then
\begin{equation}
 \Pr(\mathcal T_L\subseteq\mathcal S_{Rs})
 \geq1-L\exp\!\left(-\frac{Rsq_*}{1+(s-1)\chi q_*}\right).
 \label{eq:second-moment-reuse}
\end{equation}
To prove this, let $N_z$ count occurrences in the $s$ measurements of one circuit. Conditional independence gives
$\mathbb EN_z=s\mu_z$ and
$\mathbb EN_z^2=s\mu_z+s(s-1)\mathbb Ep_\omega(z)^2$.
Cauchy--Schwarz gives
$\Pr(N_z>0)\geq(\mathbb EN_z)^2/\mathbb EN_z^2
\geq sq_*/[1+(s-1)\chi q_*]$.
Multiply miss probabilities across the independent circuits and sum over targets. The condition on $\chi$ is additional information: $1\leq\chi_z\leq1/\mu_z$, and the upper extreme occurs when $p_\omega(z)$ takes only the values zero and one.

\section{Retained weight, reconstruction, and classical access}\label{app:mass}
\subsection{Proof of Theorem~\ref{thm:mass}}
Fix $L$ and put $\mathcal R_M=\mathcal T_L\setminus\mathcal S_M$. Every fixed subset $\mathcal R\subseteq\mathcal T_L$ is unobserved with probability
$(1-q(\mathcal R))^M\leq e^{-Mq(\mathcal R)}$.
If $q(\mathcal R_M)>u$, at least one of the at most $2^L$ subsets of $\mathcal T_L$ with mass exceeding $u$ was entirely missed. Hence
\begin{equation}
 \Pr(q(\mathcal R_M)>u)\leq2^Le^{-Mu}.
 \label{eq:subset-union}
\end{equation}
Set $u=[L\ln2+\ln(1/\delta)]/M$ and use
\begin{equation}
 1-\|P_{\mathcal S_M}x\|^2
 \leq\tau_L+\|P_{\mathcal R_M}x\|^2
 \leq\tau_L+(\epsilon+C\sqrt{q(\mathcal R_M)})^2.
\end{equation}
This proves Eq.~\eqref{eq:mass-bound}. Applying it with failure budgets $\delta/[L(L+1)]$ and summing these budgets over $1\leq L\leq N$ proves the simultaneous version. 
For reused circuits, apply the proof conditional on a good pool and replace the independent-shot calculation by Eq.~\eqref{eq:conditional-miss}. 
This establishes the extension stated after Theorem~\ref{thm:mass}.

For $M$ independent shots from $q$, an expectation bound follows from the direct occupancy calculation
\begin{equation}
 \mathbb E q(\mathcal R_M)=
 \sum_{z\in\mathcal T_L}q(z)(1-q(z))^M\leq\frac{L}{eM}.
\end{equation}
Here $(1-u)^M\leq e^{-Mu}$ and $ue^{-Mu}\leq1/(eM)$ for $u\in[0,1]$. Squaring Eq.~\eqref{eq:set-envelope} and using concavity of the square root yields
\begin{equation}
 \begin{split}
 &\mathbb E[1-\|P_{\mathcal S_M}x\|^2]\\
 &\quad\leq\min\!\left\{1,\min_{1\leq L\leq N}
       \left[\tau_L+\left(\epsilon+C\sqrt{\frac{L}{eM}}\right)^2\right]\right\}.
 \end{split}
 \label{eq:mass-expectation}
\end{equation}
The minimization is over deterministic tail bounds; no knowledge of their minimizing $L$ is required by the sampler.

If the actual independent-shot law is $q'$ with total-variation distance at most $\eta_{\mathrm{TV}}$ from $q$, apply Eq.~\eqref{eq:subset-union} to $q'$. On its success event,
$q(\mathcal R_M)\leq q'(\mathcal R_M)+\eta_{\mathrm{TV}}\leq u+\eta_{\mathrm{TV}}$.
Thus the retained-weight formula remains valid with
\begin{equation}
 \epsilon+C\sqrt{u}\quad\text{replaced by}\quad
 \epsilon+C\sqrt{u+\eta_{\mathrm{TV}}}.
 \label{eq:mass-tv}
\end{equation}
The simultaneous version uses
$u=[L\ln2+\ln(L(L+1)/\delta)]/M$.

\subsection{Proof of Proposition~\ref{prop:ritz}}
Let $Q_b=I-|b\rangle\langle b|$ and $K=Q_bA$. Invertibility gives
$\ker K=\operatorname{span}\{x\}$ and $\operatorname{rank}K=N-1$.
The nonzero eigenvalues of $K^\dagger K=H_b(A)$ equal those of $KK^\dagger=Q_bA^2Q_b$. On $b^\perp$, the latter has quadratic form between $a^2$ and $B^2$ on normalized vectors by Eq.~\eqref{eq:spectrum}. Hence its $N-1$ eigenvalues lie in that interval. Its partner $H_b(A)$ has kernel $\operatorname{span}\{x\}$, proving Eq.~\eqref{eq:sandwich}.

Write $w=\|P_{\mathcal S}x\|^2$. If $w>0$, the normalized truncation $P_{\mathcal S}x/\sqrt w$ has squared overlap $w$ with $x$. Its energy is at most $B^2(1-w)$ by the upper inequality in Eq.~\eqref{eq:sandwich}. The variational principle gives Eq.~\eqref{eq:ritz-energy}. If $w=0$, the same conclusion follows from $\|H_b(A)\|\leq B^2$. The lower inequality in Eq.~\eqref{eq:sandwich} proves Eqs.~\eqref{eq:ritz-fidelity} and \eqref{eq:certificate}.

For pure states, trace distance is the square root of infidelity. Thus a cost upper bound $e$ implies trace distance at most $\min\{1,\sqrt e/a\}$.

\subsection{Accuracy-driven and finite-precision budgets}
For a total witness budget $0<\epsilon\leq1$, first use Appendix~\ref{app:finite-choice} with a filter budget smaller than $\epsilon$, then allocate the remaining positive margin to Eq.~\eqref{eq:finite-depth}. A fixed fractional split gives
\begin{equation}
 \begin{gathered}
 C=Bdt=O(\kappa\ell),\qquad
 \ell=\Theta(1+\ln(1/\epsilon)),\\
 D=O\!\left(\left(\frac{\lambda_A}{a}\right)^2
                  \frac{\kappa\ell^3}{\epsilon}\right).
 \end{gathered}
 \label{eq:accuracy-finite-orders}
\end{equation}
This construction does not use $\beta_L$. The signed-time version has $C=B\Lambda_a=\Theta(\kappa)$ and expected length
$O(1+(\lambda_A/a)^2\kappa/\epsilon)$.

To obtain Eq.~\eqref{eq:accuracy-budget}, set
$u=[L\ln2+\ln(1/\delta)]/M$ in Theorem~\ref{thm:mass}. The budget makes
\begin{equation}
 \epsilon+C\sqrt{u}\leq\sqrt{\eta/\kappa^2-\tau_L}.
\end{equation}
After squaring and adding $\tau_L$, the missing weight in $\mathcal S_M$ is at most $\eta/\kappa^2$. Proposition~\ref{prop:ritz} therefore gives the target infidelity for exact reconstruction on that same subspace.

For the orders in Eq.~\eqref{eq:accuracy-scaling}, suppose that
$\eta/\kappa^2-\tau_L$ is at least a fixed positive fraction of $\eta/\kappa^2$. One explicit allocation is
\begin{equation}
 \epsilon=\frac12\sqrt{\eta/\kappa^2-\tau_L}.
 \label{eq:accuracy-scaling-choice}
\end{equation}
This leaves the same positive margin for sampling and gives
$\epsilon=\Theta(\sqrt\eta/\kappa)$.
Substituting $C=B\Lambda_a=\Theta(\kappa)$ into Eq.~\eqref{eq:accuracy-budget}, and this $\epsilon$ into Eq.~\eqref{eq:expected-depth}, proves Eq.~\eqref{eq:accuracy-scaling}. A supplied upper bound on $\tau_L$ may be used throughout. The finite-grid alternative adds a factor $\ell^2$ to the shot bound and $\ell^3$ to the sufficient term length, with $\ell=\Theta(1+\ln(\kappa/\sqrt\eta))$.

To prove Eq.~\eqref{eq:finite-precision}, let $\widetilde H_{\mathcal S}$ be the computed matrix and $u$ the normalized returned coefficient vector. The hypotheses imply
\begin{equation}
 \begin{split}
 u^\dagger H_{\mathcal S}u
 &\leq u^\dagger\widetilde H_{\mathcal S}u+\xi_H\\
 &\leq\lambda_{\min}(\widetilde H_{\mathcal S})+\xi_{\mathrm{eig}}+\xi_H\\
 &\leq\lambda_{\min}(H_{\mathcal S})+\xi_{\mathrm{eig}}+2\xi_H.
 \end{split}
\end{equation}
The last inequality follows by evaluating the perturbed Rayleigh quotient on a lowest eigenvector of $H_{\mathcal S}$. Combine this with Eq.~\eqref{eq:ritz-energy} and the lower sandwich bound. A computed quotient $\widetilde e=u^\dagger\widetilde H_{\mathcal S}u$ separately certifies infidelity at most $\min\{1,(\widetilde e+\xi_H)/a^2\}$. For a final target $\eta$, a sufficient condition for reconstruction on $\mathcal S_M$ is
\begin{equation}
 \begin{split}
 &\kappa^2\left[\tau_L+
 \left(\epsilon+C\sqrt{\frac{L\ln2+\ln(1/\delta)}{M}}\right)^2\right]\\
 &\quad+\frac{2\xi_H+\xi_{\mathrm{eig}}}{a^2}\leq\eta.
 \end{split}
\end{equation}
Equivalently, reserve $(2\xi_H+\xi_{\mathrm{eig}})/a^2$ from the target before applying Eqs.~\eqref{eq:accuracy-margin}--\eqref{eq:accuracy-budget}, with a positive remaining margin.

\subsection{Projected matrix construction and compact SVD}
In this subsection write $W=AE_{\mathcal S}$ and $g=W^\dagger b$. Direct multiplication gives
\begin{equation}
 H_{\mathcal S}=W^\dagger W-gg^\dagger.
 \label{eq:gram-reconstruction}
\end{equation}
This uses full selected columns of $A$, not only their entries within $\mathcal S$. Indeed,
\begin{equation}
 P_{\mathcal S}A^2P_{\mathcal S}
 =(P_{\mathcal S}AP_{\mathcal S})^2+
 P_{\mathcal S}A(I-P_{\mathcal S})AP_{\mathcal S}.
 \label{eq:projection-warning}
\end{equation}
The second term contains excursions outside the coordinate subspace and need not vanish.

If each selected column has at most $s_A$ nonzeros, constructing and sorting their row union costs, in addition to access, at most
$O(s_A|\mathcal S|\log(s_A|\mathcal S|))$ elementary comparison and bookkeeping operations. Inner products of sorted columns require $O(s_A)$ arithmetic operations per pair, giving the Gram-matrix count in the main text. The vector $g$ uses amplitude queries only on the union of those rows. Storing the sparse columns and a dense projected matrix costs
$O(s_A|\mathcal S|+|\mathcal S|^2)$ numbers.
Under full retention, $|\mathcal S_M|\leq M$. Polynomial bounds on the sampling parameters, sparse-column access, state preparation, term implementation, and precision overhead therefore give conditional polynomial resource requirements.

Subtraction in Eq.~\eqref{eq:gram-reconstruction} can lose relative accuracy at a very small Ritz energy. Let $\mathcal R$ be the row union of $W$, and restrict $W$ and $b$ to obtain $W_{\mathcal R},b_{\mathcal R}$. Set
\begin{equation}
 \overline W=\begin{pmatrix}W_{\mathcal R}\\0\end{pmatrix},\qquad
 \overline b=\begin{pmatrix}b_{\mathcal R}\\\sqrt{1-\|b_{\mathcal R}\|^2}\end{pmatrix}.
\end{equation}
Then $\|\overline b\|=1$, $\overline W^\dagger\overline W=W^\dagger W$, and $\overline W^\dagger\overline b=g$. Hence
\begin{equation}
 [(I-\overline b\,\overline b^\dagger)\overline W]^\dagger
 [(I-\overline b\,\overline b^\dagger)\overline W]=H_{\mathcal S}.
\end{equation}
A smallest right singular vector gives the desired Ritz coefficients. The compact matrix has at most $s_A|\mathcal S|+1$ rows even when $b$ is dense.

\subsection{Relation to sampled-column least squares}\label{app:least-squares}
Under Eq.~\eqref{eq:spectrum}, fix a nonempty coordinate set $\mathcal S$ and assume exact classical solves in this subsection. Since $A$ is invertible, $W=AE_{\mathcal S}$ has full column rank. Writing $G=W^\dagger W$ and $g=W^\dagger b$, least squares gives
\begin{equation}
 u_{\mathrm{LS}}=G^{-1}g,\qquad
 \min_u\|Wu-b\|^2=1-g^\dagger G^{-1}g.
\end{equation}
For a fixed normalized direction $u$, minimizing over its complex scale gives
\begin{equation}
 \min_{\gamma\in\mathbb C}\|\gamma Wu-b\|^2
 =1-\frac{|g^\dagger u|^2}{u^\dagger Gu}.
 \label{eq:scaled-ls}
\end{equation}
This follows by completing the square in $\gamma$, whose minimizing value is $(Wu)^\dagger b/\|Wu\|^2$. The Ritz objective is instead
$u^\dagger Gu-|g^\dagger u|^2$ under $\|u\|=1$.
Thus the two objectives need not have the same minimizing direction. If the selected space contains $x$, both recover its direction and least squares also determines its scale for the specified normalized right-hand side.

Put $r=b-Wu_{\mathrm{LS}}$. When $u_{\mathrm{LS}}\neq0$, define the normalized least-squares output $\widehat x_{\mathrm{LS}}=E_{\mathcal S}u_{\mathrm{LS}}/\|u_{\mathrm{LS}}\|$. Its basic residual certificate is
\begin{equation}
 1-|\langle x|\widehat x_{\mathrm{LS}}\rangle|^2
 \leq\min\{1,\kappa^2\|r\|^2\}.
 \label{eq:ls-residual-certificate}
\end{equation}
Indeed, the distance from $A^{-1}b$ to the line spanned by $E_{\mathcal S}u_{\mathrm{LS}}$ is no larger than
$\|A^{-1}b-E_{\mathcal S}u_{\mathrm{LS}}\|=\|A^{-1}r\|$.
Dividing by $\|A^{-1}b\|$, squaring, and using $\|A^{-1}\|\leq a^{-1}$ and $\|A^{-1}b\|\geq B^{-1}$ gives Eq.~\eqref{eq:ls-residual-certificate}.

The normal equations imply $(Wu_{\mathrm{LS}})^\dagger r=0$, so
$b^\dagger Wu_{\mathrm{LS}}=\|Wu_{\mathrm{LS}}\|^2=1-\|r\|^2$.
The exact Ritz energy $e_{\mathcal S}$ from Eq.~\eqref{eq:ritz-output} therefore satisfies
\begin{equation}
 \begin{aligned}
 \frac{e_{\mathcal S}}{a^2}
 &\leq\frac{\langle\widehat x_{\mathrm{LS}}|H_b(A)|\widehat x_{\mathrm{LS}}\rangle}{a^2}\\
 &=\frac{\|A\widehat x_{\mathrm{LS}}\|^2}{a^2}\,\|r\|^2\\
 &\leq\kappa^2\|r\|^2.
 \end{aligned}
 \label{eq:ritz-ls-certificate}
\end{equation}
The first inequality is the Rayleigh--Ritz minimum in the same subspace; the equality uses the preceding orthogonality, and the last inequality uses $\|A\widehat x_{\mathrm{LS}}\|\leq B$. Thus the Ritz certificate in Eq.~\eqref{eq:certificate}, after capping at one, is no larger than the residual upper bound in Eq.~\eqref{eq:ls-residual-certificate}. If $u_{\mathrm{LS}}=0$, then $g=0$, $\|r\|=1$, and $H_{\mathcal S}=G$, with $a^2\leq e_{\mathcal S}\leq B^2$. The two capped certificate values are then both one, but no normalized least-squares output is defined.

\section{Proofs of the sampling lower bounds}\label{app:lower}
\subsection{Complete support discovery}
For the first statement of Theorem~\ref{thm:lower}, put $r=L-1$. On $L$ coordinates choose
\begin{equation}
 \begin{split}
 A&=\operatorname{diag}(\underbrace{\kappa^{-1},\ldots,\kappa^{-1}}_{r},1),\\
 x&=(\underbrace{\sqrt\beta,\ldots,\sqrt\beta}_{r},\sqrt{1-r\beta})^{\mathsf T},\\
 b&=Ax/\|Ax\|.
 \end{split}
 \label{eq:lower-instance}
\end{equation}
Pad with unit diagonal entries and zero amplitudes to obtain dimension $2^n$ when needed. The assumption $r\beta\leq1/2$ ensures that the last coordinate has weight at least $\beta$, so the minimum target weight is exactly $\beta$. Every rare coordinate has probability
\begin{equation}
 p=\frac{\beta}{r\beta+\kappa^2(1-r\beta)}
 \label{eq:lower-rare-probability}
\end{equation}
in $b$ and at every real evolution time. Diagonal qDRIFT terms preserve these probabilities realization by realization. Adaptive time choices therefore cannot change the distribution of any fresh measurement.

Let $Z$ count rare coordinates unobserved after $M$ shots. Since $p\leq1/2$,
\begin{equation}
 \begin{split}
 \mathbb EZ&=r(1-p)^M,\\
 \mathbb EZ^2&=\mathbb EZ+r(r-1)(1-2p)^M\\
 &\leq\mathbb EZ+(\mathbb EZ)^2.
 \end{split}
\end{equation}
The final inequality uses $1-2p\leq(1-p)^2$. Cauchy--Schwarz on $Z\mathbf1_{Z>0}$ gives
$\Pr(Z>0)\geq\mathbb EZ/(1+\mathbb EZ)$.
Thus success at least $1-\delta$ implies
\begin{equation}
 \begin{split}
 M&\geq\frac{\ln[r(1-\delta)/\delta]}{-\ln(1-p)}\\
 &\geq\frac{\kappa^2}{2\beta}\ln\frac{r(1-\delta)}{\delta}.
 \end{split}
 \label{eq:lower-explicit}
\end{equation}
Here $-\ln(1-p)\leq p/(1-p)$ and
$(1-p)/p=r-1+\kappa^2(1-r\beta)/\beta\geq\kappa^2/(2\beta)$.
For $L\geq2$ and $\delta\leq1/4$, $r(1-\delta)/\delta\geq3L/(8\delta)$ and $L/\delta\geq8$. Hence $\ln[r(1-\delta)/\delta]\geq\tfrac12\ln(L/\delta)$, which proves Eq.~\eqref{eq:support-lower}.

\subsection{Approximate retained mass}
Keep the same diagonal matrix, but take
\begin{equation}
 x=(\underbrace{\sqrt{4\eta/r},\ldots,\sqrt{4\eta/r}}_{r},
          \sqrt{1-4\eta})^{\mathsf T},\qquad b=Ax/\|Ax\|.
\end{equation}
All $L$ coordinates are nonzero for $0<\eta\leq1/8$. The combined measurement probability of the $r$ rare coordinates is
\begin{equation}
 P_{\mathrm{rare}}=\frac{4\eta}{4\eta+\kappa^2(1-4\eta)}
 \leq\frac{8\eta}{\kappa^2}.
\end{equation}
To leave missing solution weight at most $\eta$, at least $3r/4$ distinct rare coordinates must be observed. The number of rare measurement outcomes is at least this large on the success event and has expectation $MP_{\mathrm{rare}}$. Therefore
\begin{equation}
 M\geq\frac{3(1-\delta)r}{4P_{\mathrm{rare}}}
 \geq\frac{9\kappa^2r}{128\eta}.
\end{equation}
Missing every rare coordinate also prevents success, so
$(1-P_{\mathrm{rare}})^M\leq\delta$ is necessary. It follows that
\begin{equation}
 M\geq\frac{1-P_{\mathrm{rare}}}{P_{\mathrm{rare}}}\ln\frac1\delta
 =\frac{\kappa^2(1-4\eta)}{4\eta}\ln\frac1\delta
 \geq\frac{\kappa^2}{8\eta}\ln\frac1\delta.
\end{equation}
Taking the larger of the two lower bounds proves Eq.~\eqref{eq:mass-lower}, since $r=L-1\geq L/2$. The same phase-invariance argument permits adaptive signed times and diagonal compilation.

Finally, any normalized output supported in $\mathcal S_M$ has squared overlap with $x$ at most $\|P_{\mathcal S_M}x\|^2$, by Cauchy--Schwarz. The retained-mass lower bound is therefore necessary for such an output to have infidelity at most $\eta$.

\section{Optional validation and refinement of selected supports}\label{app:selection}
The cost certificate in Eq.~\eqref{eq:certificate} applies to any computed candidate. The additional procedures here quantify support validation and coefficient compression when useful. They do not require a particular recovery rule.

\subsection{Independent validation of a proposed coordinate set}
Let a nonempty set $\mathcal B$ be proposed by any procedure, and then fixed before drawing $n_{\mathrm{val}}\geq1$ independent validation samples from a reference law $q$ satisfying Eq.~\eqref{eq:envelope}. The validation samples must be independent of the data used to select $\mathcal B$. Let $J_{\mathrm{out}}$ count outcomes outside $\mathcal B$, and set
\begin{equation}
 u_{\mathrm{val}}=
 \min\left\{1,\frac{J_{\mathrm{out}}}{n_{\mathrm{val}}}
       +\sqrt{\frac{\ln(1/\delta)}{2n_{\mathrm{val}}}}\right\}.
 \label{eq:validation-upper}
\end{equation}
With probability at least $1-\delta$,
\begin{equation}
 1-\|P_{\mathcal B}x\|^2
 \leq\min\{1,(\epsilon+C\sqrt{u_{\mathrm{val}}})^2\}.
 \label{eq:validation-mass}
\end{equation}
To prove it, condition on the proposal data. The outside indicators are independent Bernoulli variables with mean $q(\mathcal B^c)$. The one-sided Hoeffding inequality~\cite{hoeffding} gives $q(\mathcal B^c)\leq u_{\mathrm{val}}$ with the stated probability. Equation~\eqref{eq:set-envelope}, applied to $\mathcal B^c$, proves Eq.~\eqref{eq:validation-mass}. The conditional bound is uniform in the proposal and hence also holds unconditionally. Proposition~\ref{prop:ritz} then bounds the infidelity of the lowest Ritz vector in $\mathcal B$ by
\begin{equation}
 \min\{1,\kappa^2(\epsilon+C\sqrt{u_{\mathrm{val}}})^2\}.
\end{equation}

A prespecified zero-failure alternative reports
$u_0=1-\delta^{1/n_{\mathrm{val}}}$ when $J_{\mathrm{out}}=0$, and reports $1$ otherwise. Its coverage follows because a true outside probability above $u_0$ makes the all-inside event less likely than $\delta$. Also $u_0\leq\ln(1/\delta)/n_{\mathrm{val}}$. Selecting among confidence bounds after examining the validation data requires simultaneous coverage, for example by dividing the failure budget.

A finite collection of batches fixed before validation can use a common holdout set with failure budgets summing to $\delta$. An adaptive sequence can instead use fresh holdout samples at each step and preassigned budgets $\delta_j>0$ satisfying $\sum_j\delta_j\leq\delta$. 
If the actual validation law $q'$ has a certified total-variation distance at most $\eta_{\mathrm{TV}}$ from $q$, replace $u_{\mathrm{val}}$ by $\min\{1,u_{\mathrm{val}}+\eta_{\mathrm{TV}}\}$.

\subsection{Compression of a computed vector}
Let $y$ be normalized with $\langle y|H_b(A)|y\rangle\leq e$. For a coordinate set $\mathcal B$, suppose
\begin{equation}
 \sigma=\|(I-P_{\mathcal B})y\|^2<1,
 \qquad y_{\mathcal B}=P_{\mathcal B}y/\sqrt{1-\sigma}.
\end{equation}
Then
\begin{align}
 1-|\langle x|y_{\mathcal B}\rangle|^2
 &\leq\min\{1,(\sqrt e/a+\sqrt\sigma)^2\},
 \label{eq:pruning-fidelity}\\
 \langle y_{\mathcal B}|H_b(A)|y_{\mathcal B}\rangle
 &\leq\frac{(\sqrt e+B\sqrt\sigma)^2}{1-\sigma}.
 \label{eq:pruning-energy}
\end{align}
Indeed, the trace distance between the pure states of $y$ and $y_{\mathcal B}$ is $\sqrt\sigma$. The triangle inequality and Eq.~\eqref{eq:certificate} prove Eq.~\eqref{eq:pruning-fidelity}. For the energy, write $K=(I-|b\rangle\langle b|)A$, so $H_b(A)=K^\dagger K$ and $\|K\|\leq B$. Then
\begin{equation}
 \|KP_{\mathcal B}y\|\leq\|Ky\|+\|K(I-P_{\mathcal B})y\|
 \leq\sqrt e+B\sqrt\sigma.
\end{equation}
Normalization proves Eq.~\eqref{eq:pruning-energy}. If the cost Hamiltonian is rediagonalized in $\mathcal B$, its lowest Ritz vector consequently has infidelity at most
\begin{equation}
 \min\left\{1,\frac{(\sqrt e/a+\kappa\sqrt\sigma)^2}{1-\sigma}\right\}.
\end{equation}
The finite-precision correction in Eq.~\eqref{eq:finite-precision} can be added before capping at one.

The omitted weight $\sigma$ belongs to the available vector $y$, not the unknown target. Retaining its largest coefficients therefore gives a computable compression budget, provided $e$ is a reliable upper bound on the true cost.

\subsection{Cost-based acceptance}
Suppose a baseline normalized vector $y_0$ has true cost $e_0=\langle y_0|H_b(A)|y_0\rangle$. Accepting a candidate $y$ only when its true cost is at most $e_0$ preserves
\begin{equation}
 1-|\langle x|y\rangle|^2\leq\min\{1,e_0/a^2\}.
 \label{eq:acceptance-certificate}
\end{equation}
If the baseline is a Ritz solution on a space with missing weight at most $m_0$, then $e_0\leq B^2m_0$ and the original $\kappa^2m_0$ certificate is preserved. This is an immediate application of Proposition~\ref{prop:ritz}; the accepted candidate need not retain every baseline coordinate.

With inexact energy estimates, an upper confidence or deterministic bound for the candidate no greater than a corresponding lower bound for the incumbent suffices for nonincrease on the event that both bounds are valid. Repeated stochastic comparisons require simultaneous coverage, for example through an allocated failure budget.

\section{Cost-aware time sampling and compilation}\label{app:allocation}
Coefficient-weighted sampling can be adapted to unequal execution costs, as in other randomized linear-algebra constructions~\cite{wang}. The objectives below optimize the sufficient sampling bound, with the witness held fixed and a fixed positive amplitude margin between its error bound and $\sqrt{\beta_L}$.

\subsection{Unequal fixed execution costs}
For a finite witness, let $w_k\geq|c_k|$ be known positive bounds and let $G_k>0$ be the cost of one execution at time $T_k$. Terms with zero coefficient can be omitted. Holding the witness and its error fixed, the sufficient expected execution cost, before the integer ceiling on the shot budget, is proportional to
\begin{equation}
 \begin{gathered}
 F(\pi)=\left(\sum_k\frac{w_k^2}{\pi_k}\right)
             \left(\sum_k\pi_kG_k\right),\\
 \pi_k>0,\qquad\sum_k\pi_k=1.
 \end{gathered}
\end{equation}
Cauchy--Schwarz gives
\begin{equation}
 \min_\pi F(\pi)=\left(\sum_kw_k\sqrt{G_k}\right)^2,
 \qquad
 \pi_k^*=\frac{w_k/\sqrt{G_k}}{\sum_jw_j/\sqrt{G_j}}.
 \label{eq:allocation-fixed}
\end{equation}
For completeness, apply the inequality to $w_k/\sqrt{\pi_k}$ and $\sqrt{\pi_kG_k}$. Equality requires $w_k/\sqrt{\pi_k}$ to be proportional to $\sqrt{\pi_kG_k}$, which gives the displayed optimizer after normalization.

For equal costs and exact coefficient magnitudes, this is the coefficient-proportional rule following Eq.~\eqref{eq:finite-witness}. A common unknown scale in the coefficients cancels from $\pi_k^*$. Restoring the shot ceiling adds at most one mean-shot cost.

\subsection{Joint time and depth allocation}
Suppose the mean-state error at a nonzero time is bounded by $b_k/D_k$, with known $b_k>0$. For qDRIFT, $b_k=\lambda_A^2T_k^2$. Ignore fixed per-shot overhead, count one unit per term exponential, and first allow arbitrary real $D_k>0$. Choose $\epsilon_{\mathrm{av}}>0$ so that the ideal witness error plus $\epsilon_{\mathrm{av}}$ is smaller than $\sqrt{\beta_L}$, and impose
\begin{equation}
 \sum_k\frac{w_kb_k}{D_k}\leq\epsilon_{\mathrm{av}}.
 \label{eq:allocation-constraint}
\end{equation}
The relaxed cost objective is
\begin{equation}
 F(\pi,D)=\left(\sum_k\frac{w_k^2}{\pi_k}\right)
               \left(\sum_k\pi_kD_k\right).
\end{equation}
Writing $S_b=\sum_kw_kb_k^{1/3}$, its minimum and an optimizer are
\begin{align}
 \min_{\pi,D}F(\pi,D)&=\frac{S_b^3}{\epsilon_{\mathrm{av}}},
 \label{eq:allocation-joint-value}\\
 D_k^*&=\frac{S_b}{\epsilon_{\mathrm{av}}}b_k^{2/3},
 \qquad
 \pi_k^*=\frac{w_kb_k^{-1/3}}{\sum_jw_jb_j^{-1/3}}.
 \label{eq:allocation-joint}
\end{align}
To prove this, Eq.~\eqref{eq:allocation-fixed} first minimizes over $\pi$ to give $(\sum_kw_k\sqrt{D_k})^2$. H\"older's inequality yields
\begin{align}
 S_b
 &=\sum_k(w_k\sqrt{D_k})^{2/3}(w_kb_k/D_k)^{1/3}\nonumber\\
 &\leq\left(\sum_kw_k\sqrt{D_k}\right)^{2/3}
        \left(\sum_k\frac{w_kb_k}{D_k}\right)^{1/3}.
\end{align}
Cubing and using Eq.~\eqref{eq:allocation-constraint} proves the lower bound in Eq.~\eqref{eq:allocation-joint-value}. Substitution of Eq.~\eqref{eq:allocation-joint} makes the constraint tight and attains equality in both inequalities.

In this qDRIFT surrogate, $D_k^*\propto|T_k|^{4/3}$ and $\pi_k^*\propto w_k|T_k|^{-2/3}$. 
This differs from assigning depths proportional to $T_k^2$ before optimizing the time law. 
Restoring integer lengths by $D_k^{\mathrm{int}}=\max\{1,\lceil D_k^*\rceil\}$ preserves the error bound and, with the same $\pi^*$, increases the objective by at most $\sum_kw_k^2/\pi_k^*$. 
Reoptimizing the probabilities for the rounded lengths can only reduce that rounded upper bound.

\makeatletter
\renewcommand{\bibsection}{%
  \par
  \begingroup
    \baselineskip26\p@
    \bib@device{\hsize}{72\p@}%
  \endgroup
  \nobreak\@nobreaktrue
  \addvspace{19\p@}%
}
\makeatother

\bibliography{references}

\end{document}